\documentclass[iop,apj,tighten]{emulateapj}

\usepackage{amsmath,amstext}
\usepackage[breaklinks]{hyperref} 
\usepackage[hyphenbreaks]{breakurl}
\usepackage{color}

\newcommand\Msun{{\,M_\odot}}
\slugcomment{submitted to the Astrophysical Journal}
\shorttitle{The Frontier Fields}
\shortauthors{Lotz et al. }

\begin{document}

\title{The Frontier Fields: Survey Design}

\author{J. M. Lotz\altaffilmark{1,2}, A. Koekemoer\altaffilmark{1},  D. Coe\altaffilmark{3}, N. Grogin\altaffilmark{1}
  P. Capak\altaffilmark{3},    J. Mack\altaffilmark{1},  J. Anderson\altaffilmark{1},  R. Avila\altaffilmark{1}, E. A. Barker\altaffilmark{1}, 
D. Borncamp\altaffilmark{1}, G. Brammer\altaffilmark{3}, M. Durbin\altaffilmark{1}, H. Gunning\altaffilmark{1}, 
B. Hilbert\altaffilmark{1}, H. Jenkner\altaffilmark{1}, H. Khandrika\altaffilmark{1}, Z. Levay\altaffilmark{1}
R. A. Lucas\altaffilmark{1},  J. MacKenty\altaffilmark{1}, S. Ogaz\altaffilmark{1},  B. Porterfield\altaffilmark{1},  N. Reid\altaffilmark{1}, 
M. Robberto\altaffilmark{1}, P. Royle\altaffilmark{1}, L. J. Smith\altaffilmark{3},  L. J. Storrie-Lombardi\altaffilmark{4},  B. Sunnquist\altaffilmark{1}, J. Surace\altaffilmark{4}  D. C. Taylor\altaffilmark{1},     R. Williams\altaffilmark{1},   
J. Bullock\altaffilmark{5}, M. Dickinson\altaffilmark{6}, S. Finkelstein\altaffilmark{7},  P. Natarajan\altaffilmark{8},  J. Richard\altaffilmark{9}, B. Robertson\altaffilmark{10}, J. Tumlinson\altaffilmark{1},  A. Zitrin\altaffilmark{11}, 
K. Flanagan\altaffilmark{1},  K. Sembach\altaffilmark{1}, B. T. Soifer\altaffilmark{3} \\ and \\ M. Mountain\altaffilmark{1}}

\altaffiltext{1}{Space Telescope Science Institute, 3700 San Martin  Dr., Baltimore, MD 21218, USA}
\altaffiltext{2}{email: lotz@stsci.edu}
\altaffiltext{3}{European Space Agency/Space Telescope Science Institute}
\altaffiltext{4}{Spitzer Science Center, California Institute of Technology}
\altaffiltext{5}{Physics \& Astronomy Department, University of California, Irvine, CA}
\altaffiltext{6}{National Optical Astronomy Observatory, Tucson, AZ }
\altaffiltext{7}{Department of Astronomy, University of Texas, Austin, TX}
\altaffiltext{8}{Department of Astronomy, Yale University,  New Haven, CT}
\altaffiltext{9}{Univ Lyon, Univ Lyon1, Ens de Lyon, CNRS, Centre de Recherche Astrophysique de Lyon UMR5574, F-69230, Saint-Genis-Laval, France}
\altaffiltext{10}{Department of Astronomy \& Astrophysics, University of California,  Santa Cruz, CA }
\altaffiltext{11}{Department of Astronomy, California Institute of Technology, Pasadena, CA}

\begin{abstract}

How deep can we go?    What are the faintest and most distant galaxies we can see with the {\it Hubble Space Telescope} now, 
before the launch of the {\it James Webb Space Telescope}?   This is the challenge taken up by the Frontier Fields,  
a director's discretionary time campaign with {\it HST} and the {\it Spitzer Space Telescope} to see deeper into the 
universe than ever before. The Frontier Fields combines the power of {\it HST} and {\it Spitzer} with the natural gravitational 
telescopes of massive high-magnification clusters of galaxies to produce the deepest observations of clusters and their lensed galaxies 
ever obtained. Six clusters $-$ Abell 2744, MACSJ0416.1-2403, MACSJ0717.5+3745, 
MACSJ1149.5+2223,  Abell S1063, and Abell 370 $-$ were selected based on their lensing strength, sky darkness, Galactic extinction, parallel field suitability, accessibility to ground-based facilities,  {\it HST, Spitzer} and {\it JWST} observability, and pre-existing ancillary data.  
These clusters have been targeted by the {\it HST ACS/WFC} and {\it WFC3/IR} cameras with coordinated parallels of adjacent blank fields 
for over 840 {\it HST} orbits in Cycles 22, 23, and 24.   The {\it Spitzer Space Telescope} has dedicated $>$ 1000 hours of 
director's discretionary time to obtain {\it IRAC} 3.6 and 4.5 micron imaging to $\sim$ 26.5, 26.0 ABmag 5$\sigma$ point-source depths in the six cluster and 
six parallel Frontier Fields. The Frontier Field parallel fields are the second-deepest observations thus far by {\it HST} with  $\sim$ 29th ABmag 5$\sigma$ point source depths in seven optical $-$ near-infrared bandpasses.  
Galaxies behind the Frontier Field cluster lenses experience typical magnification 
factors of a few,  with small regions near the critical curves magnified by factors $10-100$.  Therefore, the Frontier Field cluster {\it HST} images achieve {\it intrinsic} depths of $\sim 30-33$ magnitudes over very small volumes. Early studies of the Frontier Fields have probed galaxies 
fainter than any seen before during the epoch of reionization $6 < z < 10$,  mapped out the cluster dark matter to unprecedented resolution, and followed lensed transient events. 
The Frontier Fields DD data and public lensing models are non-proprietary and available at the {\it Mikulski Archive for Space Telescopes} and the {\it NASA/IPAC Infrared Science Archive}.  
\end{abstract}

\keywords{galaxies:clusters; galaxies:high-redshift}

\section{Introduction}
Exceptionally deep observations of the distant universe with 
the Hubble Space Telescope ({\it HST}) have consistently pushed the frontiers of human knowledge.
A succession of observing programs with each generation of {\it HST} detectors, 
in concert with the other {\it NASA} Great Observatories ({\it Spitzer Space Telescope} and {\it Chandra X-ray Observatory}),  
have probed the star-formation and assembly histories of galaxies through $>$ 95\% of the universe's lifetime.   
These observations have been made publicly available to the greater astronomy community,  enabling a wide range
of science and ancillary observing programs.   The study of {\it HST} deep fields has established a number of 
techniques now standard in extra-galactic astronomy, including the Lyman break selection of distant star-forming galaxies; 
photometric redshift determinations;  stellar population fitting to multi-band photometry;  quantitative morphological 
analysis; and the detection of high-redshift transient phenomena.    
Here we present the new Frontier Fields,  an {\it HST} and {\it Spitzer} director's discretionary 
time campaign to observe six massive strong-lensing clusters and six parallel fields,  
designed to simultaneously detect the faintest galaxies ever 
observed and provide a statistical picture of galaxy evolution at early times. 

The first Hubble Deep Field (HDF) observations with {\it HST} {\it WFPC2} revealed thousands of galaxies
to 30th magnitude, fainter than any seen before (Williams et al. 1996; Ferguson, Dickinson, \& Williams 2000).   Utilizing the
Lyman break technique (Songalia, Cowie, \& Lilly 1990;  Guhathakurta, Tyson, \& Majewski 1990),  
the HDF and subsequent HDF-South (HDF-S; Castertano et al. 2000; Williams et al. 2000; Ferguson, Dickinson, \& Williams 2000) 
detected significant numbers of distant star-forming galaxies visible in optical out to redshifts $z \sim 5$  (e.g. Madau et al. 1996). 
{\it HST}'s deep and high spatial resolution images showed that many of these distant galaxies were smaller  
with higher surface-brightnesses and more irregular structures than local galaxy populations (e.g. Abraham et al. 1996). 

Follow-up observations of the HDF and HDF-S in the infrared with {\it HST}'s {\it NICMOS} camera 
(Dickinson 1999; Thompson et al. 1999;  Franx 2003) enabled 
studies of the stellar mass of the $z < 5$ populations (e.g. Papovich et al. 2001; Dickinson et al.2003; Fontana et al. 2003)
as well as the detection of higher redshift galaxies at $5 < z < 7$ (Thompson 2003; Bouwens et al. 2003) 
and intrinsically redder populations (Labbe et al. 2003; Fern{\'a}ndez-Soto, Lanzetta, \& Yahil 1999, Stiavelli et al. 1999).   
Combined with the spectroscopic confirmation of many of these faint galaxies (e.g. Lowenthal et al. 1997; Steidel et al. 1996),  
it became possible to track the cosmic star-formation  (Madau et al. 1996; Lanzetta et al.2002; Bouwens et al. 2003)
 and assembly history of stellar mass (Dickinson et al. 2003)  over the majority of the universe's lifetime.  
{\it HST} {\it NICMOS} observations of the HDF in 1997 discovered the highest redshift Type Ia supernova known at that time ($z=1.7$),  
confirming the acceleration of the universe (Riess et al. 2001). 

After the original HDFs, synergistic multi-wavelength deep observations with Great Observatories and new capabilities on Hubble 
further expanded the boundaries of our understanding.  The installation of the Advanced Camera for Surveys Wide Field Camera ({\it ACS/WFC}; Ford et al. 1998)
on {\it HST} in 2002 greatly improved the depth and area of optical imaging possible within a fixed exposure time.  The fields for the Great Observatories Origins Deep Survey
 (GOODS; Giavalisco et al. 2004) were chosen to overlap with existing X-ray deep fields from {\it Chandra} (HDF/Chandra Deep 
Field North and the new Chandra Deep Field South; Hornschemeier et al. 2000; Giacconi et al. 2001).  New {\it HST} and 
{\it Spitzer} imaging produced high-quality and deep multi-wavelength photometry, revealed new distant galaxy populations, 
measured photometric redshifts,  improved stellar mass estimates,  and could be matched to faint X-ray sources in the 
Chandra Deep Fields (e.g. Pope et al. 2006; Mosbasher et al. 2004; Grazian et al. 2006; Fontana et al. 2006; Treister et al. 2004; Barger et al. 2005). 
 The cadence of the {\it HST} GOODS observations were designed to perform a systematic search for high-redshift supernovae (Riess et al. 2004 a, b).
The {\it HST} Ultra Deep Field (HUDF; Beckwith et al. 2006) location within GOODS-S/CDFS was chosen to leverage this existing data  
 with an additional 400 orbits (268 hours) to reach optical depths fainter than original HDF {\it WFPC2} limits.  
The resulting ``wedding cake'' survey of the combined GOODS and HUDF observations 
proved to be an important strategy for spanning the depth and area needed to 
constrain both the bright and faint ends of the luminosity function of galaxies approaching the epoch of reionization  (e.g. Bouwens et al.2007). 

With the success of the {\it HST} SM4 in 2009 and the installation of the Wide Field Camera 3 ({\it WFC3}; MacKenty et al.2008) with its 
IR channel,  {\it HST} greatly improved the efficiency of its high-spatial resolution near-infrared imaging.  
The {\it WFC3} Early Release Science near-infrared observations of GOODS-S (Windhorst et al. 2011) and deep imaging in HUDF  and parallels revealed 
new populations of galaxies at $z \sim s8$  (Illingworth \& Bouwens 2010; Bouwens et al. 2010). 
Additional {\it WFC3} observations of the HUDF (Ellis et al. 2013) added the F140W filter and deeper observations in F105W and F160W filters 
to increase the detection efficiency of highest redshift candidates ($8.5 < z < 12$).      (See also Illingworth et al.2013 for a separate reduction of all HUDF data). Wider field near-infrared imaging with the {\it HST} Multi-Cycle Treasury Cosmic Assembly Near-infrared Deep Extra-galactic Legacy Survey (CANDELS; Grogin et al. 2012; Koekemoer et al. 2012)  
built upon the previous {\it HST} ACS/WFC and {\it Spitzer} observations of the GOODS, GEMS (Rix et al. 2004), COSMOS (Scoville et al. 2007), EGS (Davis et al. 2007), and UDS 
(Lawrence et al. 2007) extragalactic legacy fields.  Thanks to {\it WFC3},  detections of $z \sim 8$ candidates are
 now relatively commonplace (e.g. Labbe et al. 2010;  Finkelstein et al. 2010; Yan et al. 2011;  McClure et al. 2011, Bradley et al. 2012).   
The current measurement of the cosmic star-formation history extends to less than 500 Myr after the 
Big Bang (e.g. Ellis et al. 2013;  Finkelstein et al. 2015; Oesch et al. 2013;  Oesch et al. 2016; but see Pirzkal et al. 2013, 
Brammer et al. 2013),  albeit with very small numbers of candidates at $z > 9$.   
{\it HST}'s observations of high redshift galaxies have placed important constraints on cosmological measures of reionization  (e.g. Robertson et al. 2015, Finkelstein et al. 2015).  

With the launch of the {\it James Webb Space Telescope} ({\it JWST}) still several years away,  and no new servicing missions to {\it HST} planned,  
significant progress on understanding the first billion years of the universe with the remaining {\it HST} years poses a major challenge. The {\it HST} and {\it Spitzer} projects proposed supporting a new joint ‘Deep Fields’ program supported with director’s discretionary time in their 2012 NASA Senior Review proposals. The Hubble Deep Fields Initiative science working group (HDFI SWG) was convened by {\it STScI} Director M. Mountain in 2012.  They
recommended a new strategy to ``go deep'': use massive clusters of galaxies 
as cosmic telescopes, combined with very deep {\it HST} and {\it Spitzer} observations\footnote{\url{www.stsci.edu/hst/campaigns/frontier-fields/documents/HDFI_SWGReport2012.pdf}}.   
Very massive clusters of galaxies are the most massive structures in the universe,  bending space-time to 
create efficient gravitational lenses (e.g. Kneib \& Natarajan 2011).  The light from galaxies behind these natural telescopes experience magnification factors of a few
 within a few arc-minutes of the cluster cores,  and magnifications $\sim$10 or greater within smaller windows along 
the critical curves.    Therefore, {\it HST} observations of these strongly-lensed fields can probe galaxies as intrinsically faint
or fainter than those detected in the HUDF in a much shorter exposure time  -- provided those galaxies fall within the
high magnification windows.   

The advantages of this strategy had already been demonstrated by
the Cluster Lensing and Supernova Survey (CLASH; Postman et al. 2012), a 524-orbit {\it HST} Multi-Cycle Treasury Program to study the gravitational lensing properties of 25 galaxy clusters.
CLASH targeted each cluster with shallow observations in 16 ultra-violet -- near-infrared {\it HST} bandpasses, in order to obtain precise photometric redshift constraints on background lensed galaxies. Within only a few orbits of {\it HST} time in the reddest filters, CLASH discovered several $z>9$ galaxy candidates highly magnified by intervening massive clusters at $z \sim 0.5$  (Coe et al. 2013;  Zheng et al. 2012; Bouwens et al. 2014).

The Frontier Fields program is an ambitious multi-cycle director's discretionary time observing campaign with 
{\it HST} and {\it Spitzer Space Telescope}  to peer deeper into the universe than ever before.  The Frontier Fields combine
 the power of {\it HST} with the natural gravitational telescopes of six high-magnification clusters of galaxies to 
 produce the deepest observations of clusters and their lensed galaxies ever obtained.   The {\it HST} cluster images are 
obtained in parallel with six parallel `blank' field images;  the
parallel field images are the second deepest images ever obtained,  and triple the blank field area imaged to 29th ABmag depths.  The  {\it Spitzer Space Telescope}
is also dedicating $ >1000$ hours of Director's discretionary time to obtain $IRAC$ 3.6 and 4.5 micron imaging to 26.5, 26.0 ABmag depths
in the six cluster and six parallel Frontier Fields.  In this paper, we describe the primary science goals in \S2;  the field selection criterion in \S3; 
the Frontier Field clusters and parallel fields in \S4; the {\it HST} and {\it Spitzer} observations in \S5;   and the public Frontier
Fields lensing modeling effort in \S6.  
Further details,  the latest {\it HST} data releases,  and Frontier Fields updates
may be found at \url{www.stsci.edu/hst/campaigns/frontier-fields/} . Details describing 
the {\it Spitzer} observations will be presented in Capak et al. 2016 (in prep) and more information is available at \url{ssc.spitzer.caltech.edu/warmmission/scheduling/approvedprograms/ddt/frontier/}.

\section{Science Goals \& Strategy}
The primary science goals of the Frontier Fields  are to explore the 
high-redshift universe accessible only with deep {\it HST} observations,  and to set the scene for {\it JWST} studies 
of the early universe.  High-redshift quasar absorption lines studies have 
found that the epoch of reionization was completed by $z \sim 6$ (Fan et al. 2006),  
while cosmic microwave background observations place the start  of reionization before $z \sim 10$ 
(e.g. Spergel et al.2003, Hinshaw et al. 2013, Planck Collaboration 2015).  
Including recent estimates of the optical depth from PLANCK data, the era between $z \sim 11$ and $z \sim 6$ probed by the deepest and reddest {\it HST} observations marks
a critical transition in the universe's history (e.g. Planck Collaboration 2015, Robertson et al. 2015).   

The installation of the {\it HST} {\it WFC3} camera with 
the near-infrared channel dramatically increased the number of galaxy candidates detected  at $z > 6$.   
However,  prior to the start of the Frontier Fields in 2013,  astronomers' understanding
of  the galaxy populations during the epoch of reionization were based largely on those detected in direct 
{\it HST} {\it WFC3/IR} imaging surveys (HUDF, CANDELS, BORG) 
and handfuls of lensed objects in shallow {\it HST} observations from CLASH.  The detected 
unlensed galaxies are the most luminous objects of their era,  and thus  significantly more massive and rare than the progenitors of today's Milky Way galaxies (e.g. Behroozi, Conroy, \& Wechsler 2013;
Boylan-Kolchin, Bullock, \& Garrison-Kimmel 2014).   High redshift galaxies are barely resolved by {\it HST}
(Oesch et al. 2010,  Ono et al. 2013),  with lensed $z >8$ galaxies  yielding intrinsic sizes less than a few hundred pcs 
across (Coe et al. 2013).  Because such high-redshift galaxies are often only observable 
in the reddest {\it HST} bandpasses,  limited 
information about their rest-frame ultraviolet slopes, stellar populations, and dust content can be inferred from their observed colors  (e.g. Finkelstein et al. 2012). 
Unseen $z > 6$ dwarf galaxies well below {\it HST}'s nominal direct detection limit are needed
to produce the number of ionizing photons required to  disassociate the universe's reservoir of intergalactic neutral hydrogen 
(e.g. Finkelstein et al.2015,   Robertson et al. 2015).   
Very few candidates at $z \sim 9$ and above were identified (Ellis et al. 2013;  Oesch et al. 2013; Zheng et al. 2012; Coe et al. 2013),  resulting in 
vigorous debate about how quickly the first star-formation proceeded and how many $z > 9$ objects future JWST might see (Oesch et al. 2012).  (The role of early black holes in terms of their contribution to the reionization budget is presently unknown and this will be revealed by JWST.)

In order to address many of these unknowns,  the Frontier Fields program was designed with the following science aims: 

1. To reveal populations $z=5-10$ galaxies that are $>10$ times fainter than any presently
known, the key building blocks of  $\sim L^*$ galaxies in the local universe.

2. To characterize the stellar populations of faint galaxies at high redshift and solidify
our understanding of the stellar mass function at the earliest times.

3. To provide, for the first time, a statistical morphological characterization of star forming
galaxies at $z > 5$.

4. To find $z > 8$ galaxies stretched out enough by foreground clusters to measure sizes
and internal structure and/or magnified enough for spectroscopic follow up.

The Frontier Fields combines several previous high-redshift galaxy observing strategies to achieve these aims:   
very deep multiband {\it HST} imaging to identify very faint distant galaxy candidates by their color; 
and  strong-gravitational lensing by massive clusters of galaxies to probe galaxies fainter than those accessible with 
direct `blank' field {\it HST} imaging.  Deep imaging with the {\it Spitzer} $IRAC$ 3.6 and 4.5 micron bands are also
required to improve photometric redshifts,  measure stellar masses and specific star-formation rates, 
and rule out low-redshift interlopers (e.g. Labb\'{e} et al. 2013; Brad\u{a}c et al. 2014).   
The clusters and their exact pointings were selected to optimize the number of
detectable $z \sim 10$ objects within the {\it HST} {\it WFC3}/$IR$ field of view  
magnified by factors of $\sim 1.5-100$, depending on their positions relative to the
critical curves of the clusters.  The {\it HST} exposure times were chosen to  probe intrinsic depths
$> 10 \times$ fainter than the HUDF in the highest magnification regions of the lensed fields, but with significantly less
time than blank field observations.  The volumes probed at the highest magnifications are 
very small (see Coe, Bradley, \& Zitrin 2015), thus the program observes multiple
clusters to improve the statistical likelihood of capturing the light from the faintest 
and most distant galaxies.    While color, redshift,  and other relative measures such
as specific star-formation rates and emission-line equivalent widths are immune to 
errors in the magnification estimates,  measurements of the intrinsic luminosities and sizes of individual objects depend 
directly on the inferred lensing magnifications.  (Integrated quantities such as galaxy luminosity functions are less susceptible to magnification uncertainties.)  In concert with the DD observing campaigns,  a unified effort to create high fidelity
public maps of the lensing properties of each FF cluster is an integral part of the FF (see \S6). 

Because each cluster is observed at a fixed {\it HST} roll angle for an extended period, we also obtain simultaneous deep parallel field observations at a single pointing centered $\sim$ 6 arcmins from the cluster core ( $>$ 1.8 projected co-moving Mpc for a $z > 0.3$ lensing cluster).  
These six new 'blank fields' are comparable in depth to the HUDF parallel fields (Oesch et al. 2007),  
and triple the area of unlensed fields observed by {\it HST} to depths $\sim$ 29th magnitude ABmag. 
The background volumes lensed by the clusters are much smaller than those probed by unlensed fields.   Thus,  while the cluster pointings
allow us to see intrinsically fainter objects than the HUDF within small volumes, the parallel fields provide a dramatic 
improvement in the volume and statistical counting of distant galaxies brighter than 29th magnitude.   This is particularly 
important for understanding the biases associated with cosmic variance  - i.e. the fact that every single sightline through the universe is unique (e.g. Robertson et al. 2014).

The Frontier Fields will set the stage for the {\it James Webb Space Telescope} to study first light galaxies at $z > 10$ and to understand the
assembly of galaxies over cosmic time.  {\it JWST} is a 6.5m cold telescope sensitive at $0.7-27$ microns, to be launched 
at the end of 2018 with a limited lifetime requirement of 5 years and a goal of 10 years.  Because {\it JWST}'s  lifetime is short relative to {\it HST},  
it is important for the astronomical community to be prepared for {\it JWST} observations early on. 
The high-redshift galaxy candidates detected by the Frontier Fields are likely to be among the first spectroscopic targets for {\it JWST},  and current studies 
will produce a better understanding of  the high-redshift galaxy luminosity functions, spectral-energy distributions, 
and sizes needed to effectively plan for {\it JWST} surveys.   The {\it HST} Frontier Fields high-resolution optical imaging shortward of 0.7 micron in $ACS$ 
F435W and F606W  reaches depths comparable to those achieved by {\it JWST} {\it NIRCam} within 1--2 hours, and hence provides an important legacy dataset for future {\it JWST} extragalactic work. 
Finally,  direct observations of the faintest first galaxies and the dwarf galaxies and early accreting black holes expected to be responsible for reionization will be challenging even with {\it JWST}.  
Development of cluster lens modeling techniques now will enable future {\it JWST} studies of strong-lensing clusters and their lensed galaxies. 

The Frontier Fields data offers the opportunity to do ground-breaking science in a number of fields other than the highest redshift universe.  Several complementary {\it HST} GO observing programs have been awarded to obtain deep WFC3/UV imaging (GO 13389, 14209; B. Siana), WFC3/IR grism spectroscopy (GO 13459; T. Treu), and target-of-opportunity follow-up of transient events (GO 13386, 13790, 14208; S. Rodney). 
Hundreds of multiply-imaged background galaxies at all redshifts have permitted the construction of dark matter maps of the clusters at unprecedented
resolution to probe cluster substructure (e.g. Jauzac et al. 2014, 2015; Wang et al. 2015, Hoag et al. 2016, Limousin, M. et al. 2016; Mohammed et al. 2016, Natarayan in prep), 
and will enable new cosmological constraints via angular scaling relations (e.g. Kneib \& Natarayan 2011). At the recommendation of the HFF review committee, an exercise comparing the various independent lens modeling methodologies and their fidelity has been on-going and the first results where more than 10 independent research groups participated in prepartion (Meneghetti et al. 2016).
Detailed studies of intermediate redshift galaxies observed both at high magnification and in deep parallel imaging will probe their internal structures,  stellar populations, and luminosity functions (e.g. Alavi et al. 2014;  Jones et al. 2015; Livermore et al. 2012; Castellano et al. 2016; Pope et al. 2016)
These deepest-ever images of massive galaxy clusters have detected intracluster light, ram-pressure stripping and tidal streams at $z > 0.3$ (e.g. Montes \& Trujillo 2014; McPartland et al. 2016),  probing the dynamic processes
 impacting galaxy evolution within these unique environments.  The new {\it HST} Frontier Fields observations have detected a number of transients (e.g. Rodney et al. 2015), including the light-curves from the first multiply-imaged supernova (Kelly et al. 2015; discovered in GLASS).  

\section{Field Selection}

The six Frontier Field clusters and parallel fields (Table \ref{table1}) were selected to meet the primary scientific goals outlined in the HDFI SWG
recommendations,  as well as to optimize the {\it HST} and {\it Spitzer} observing campaigns.   A list of 25 cluster
candidates were suggested by the HDFI SWG, and
additional candidates were suggested by the community during the
selection process.  Each cluster was evaluated using the following 
criteria. 

{\it Lensing properties:}  The primary consideration for selecting each
of the Frontier Fields was the lensing strength of the cluster. 
Each cluster's lensing strength was evaluated by
calculating the likelihood of observing a $z=9.6$ galaxy magnified to 
$H_{F160W}$ $\leq$ 27 ABmag within the {\it HST WFC3/IR} field of view, ignoring
corrections for incompleteness or sky brightness (Table 2).  Preliminary lensing models
were provided by two independent modelers,  J. Richard and A.
Zitrin, and lensing probabilities were calculated assuming a luminosity function with $\phi^{*} = 4.27 \times 10^{-4}$, $M_{UV}^{*} = -19.5$, and $\alpha = -1.98$,  
extrapolated from $z \sim 8$ (Bradley et al. 2012) by assuming $dM^{*}/dz = 0.46$ (Coe et al. 2015).  
We excluded several lower-redshift $z<0.3$ strong-lensing clusters (e.g. Abell
1689) because we could not adequately sample the low-redshift cluster
critical curves within a single {\it WFC3/IR} $2'.2 \times 2.0$ pointing.   However, although
the $z=0.308$ merging cluster Abell 2744's critical curves could not be
covered by a single {\it WFC3/IR} pointing,  the probability of observing a
$z=9.6$ galaxy near its core was among the highest of all cluster
candidates.  Because we based our selection upon the results of the
lensing model predictions,  our selection was biased towards better studied
clusters with existing imaging and spectroscopic data from which
lensing models could be
constructed.    Some otherwise promising
clusters (e.g. El Gordo; Menanteau et al. 2012) could not be evaluated as insufficient lensing model constraints were available at the time of selection. 

{\it Sky brightness and Galactic extinction:}  Observations of the very
faint extra-galactic universe are limited by the brightness of the sky
and by foreground Galactic extinction.   Zodiacal light
can have a significant impact on the depths obtained by {\it HST} and {\it Spitzer}
imaging within a given exposure time.   This background depends upon the
angular distance of the target from the Sun and the ecliptic.  Targets observed 
with high zodiacal backgrounds have near-infrared sky brightnesses several magnitudes 
brighter than the lowest zodiacal backgrounds,  resulting in significantly lower signal-to-noise
images within a given exposure time. Given the highly constrained roll-angles required 
to obtain observations of a fixed parallel field with both the {\it WFC3} and {\it ACS} cameras, 
we have a limited ability to mitigate the impact of the zodiacal background 
by constraining the solar avoidance angle. Therefore,  strong
preference was given to clusters at high ecliptic latitudes.   This
selection criteria excluded a number of strong-lensing clusters at low
ecliptic latitudes.   Additionally,  clusters at high
Galactic latitude with low extinction were strongly preferred.  MACS0717.5+3745
has relatively high Galactic extinction,  with $E_{B-V} = 0.068$ (Schlegel \& Finkbeiner 2011). 
However, this cluster was the second strongest potential lenser on our list
of candidates (Table 2). Estimates of the $H_{F160W}$ zodiacal background at the epoch of observation 
and Galactic extinction for each cluster are given in Table 1. 

\begin{figure*}
\plotone{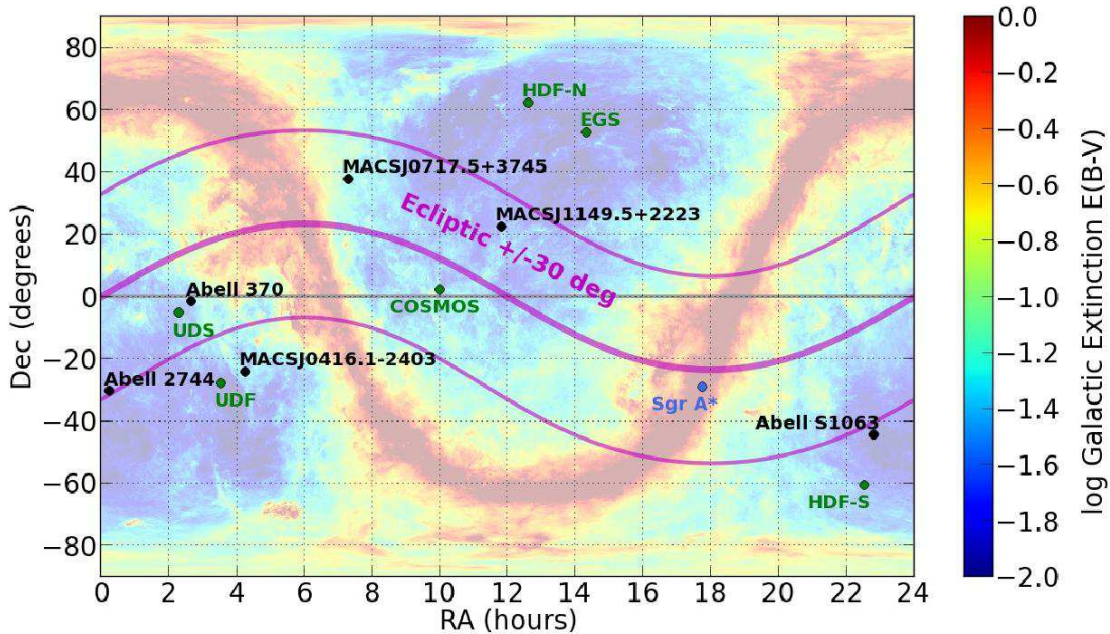}
\caption{The location of the six Frontier Field cluster $+$ parallel field pairs, 
relative to the ecliptic and Galactic plane.   The Galactic extinction map is from Schlagel, 
Finkbeiner, \& Davis (1998).   Deep extra-galactic legacy fields HDF-N, HDF-S, UDF, COSMOS,
EGS, and UDS are shown for reference. }
\end{figure*}

{\it Suitability of available parallel fields:}  The {\it HST} observing
strategy requires the simultaneous observation of the cluster field
and a blank parallel field with {\it WFC3/IR} and {\it ACS} cameras.  As we discuss
below, this observing requirement limits the range of available roll
angles, and hence locations for the parallel fields.   The potential parallel
field locations were selected to avoid bright stars and extended
cluster structures when possible.    The weak lensing signal for
each of the parallel fields was also examined where possible  
(private communication, J. Merten,  E. Medezinski, K. Umetsu).  The weak lensing signal within the parallel fields has median magnification factors between 1.02 and 1.30 for background
galaxies between $1 < z < 9$;  see discussion of each cluster for detailed estimates.

{\it Suitability for ground-based follow-up:}   Follow-up of
interesting objects detected in the Frontier Fields requires
access to those fields from the major ground-based facilities.
{\it ALMA} in particular has the potential to spectroscopically confirm the
redshift of very high redshift ($z >  6$) galaxy candidates via the
[CII] 158 micron and other atomic emission lines (e.g. da Cunha et al. 2013).   
Additionally, spectroscopic redshifts of
multiply imaged galaxies add strong constraints to the lensing models
for the clusters.   Thus,  access to the telescopes on Maunakea, in
addition to southern facilities like {\it ALMA} and {\it VLT},  were a major
consideration.   Five out of the six selected clusters are visible from
{\it ALMA}, with MACS0717.5+3745 as the exception (Tables 1, 2).   Five out of the six clusters are visible from Maunakea, with Abell S1063 as the exception.

{\it Existing ancillary data:}  Supporting data was a key
consideration recommended by the HDFI SWG.   Many of the candidate
clusters have been studied previously by space missions, including {\it HST}, the {\it Spitzer} cryo-mission with {\it  MIPS} and {\it IRAC} (including 5, 8 micron channels); 
{\it Herschel}, {\it XMM}, and {\it Chandra} (see the discussion of each cluster for details). Additionally,  ground-based spectroscopic and wide-field imaging
survey data  were evaluated from the literature.   Four of the chosen
clusters were drawn from the CLASH survey  (Postman et al. 2012),  with supporting
multi-band shallow {\it HST} imaging,  wide-field ground-based imaging ({\it Subaru}),
spectroscopy ({\it VLT}),  as well as archival {\it Herschel} and {\it Chandra} data.   
Since the announcement of the Frontier Field selection,  the community
has responded with additional observations with {\it Chandra} (PI S. Murray; C. Jones-Forman),
{\it VLA} (PI E. Murphy), {\it XMM} (PI J.P. Kneib, Eckert et al. 2015), {\it ALMA} (PI F. Bauer),  {\it LMT} (PI A. Pope), {\it Gemini} GeMS/GSAOI $K_s$ imaging (e.g. Schirmer et al. 2014),  {\it VLT} Hawk-I $K_s$ imaging (PI D. Marchesini \& G. Brammer), {\it VLT} MUSE spectroscopy (PIs Caputi \& Cl\'{e}ment, Bauer, Richard, Grillo, e.g. Karman et al. 2015, Grillo et al. 2016), as well as the release of previously unpublished data on these fields (e.g. Ebeling et al. 2014; Gruen et al. 2014).  We continue to maintain clearing-house website for public data links
and Frontier Fields-related publications: 
\url{www.stsci.edu/hst/campaigns/frontier-fields/FF-Data}. 

In addition to the science-driven considerations given above,  we
optimized the cluster selection for a number of practical issues. 

{\it HST observability:}   The Frontier Fields are observed with {\it HST} at
a fixed roll angle and its 180 degree offset in order to obtain deep observations in the cluster
field and parallel field with both {\it WFC3/IR} and ACS.   These
observations are 70 orbits at each orientation.   Each field was evaluated
to determine the ability to hold a fixed roll angle for more than 30
days and the availability of guide stars at these orientations.  For optimal
stability,  {\it HST} requires two guide stars with magnitudes brighter than
15th magnitude.  Our initial evaluation of MACSJ1149.5+2223 found only 
one acceptable guide star;  however, a second guide star with a magnitude slightly fainter than the nominal limit was available.    This new guide star was tested in early observations and found to be suitable. 

{\it Spitzer observability: }  Each cluster and parallel field was
evaluated by the {\it Spitzer} implementation team.    {\it Spitzer} observations
are sensitive to bright stars in the field,  as saturation above $\sim$35,000 DN 
can result in ``column pull-down'' impacting the data quality along the effected column.  
MACSJ0647.7+7015 (e.g. Coe et al. 2013) 
in particular was found to have unacceptably bright stars in the vicinity, and was excluded. 

{\it Schedulability: }   Each set of cluster/parallel field observations
constitutes a considerable investment of {\it HST} time,  with 70 orbits at
each orient and 140 orbits total per field.    The optimal scheduling
of these observations is a challenge.   We also anticipated that the
Frontier Fields would be popular fields for ancillary {\it HST} observing
programs.   Therefore to avoid schedule collisions with the main
Frontier Field program,  supporting Frontier Field programs, and other
popular {\it HST} fields (e.g. the UDF/GOODS-S),   the Frontier Fields were
selected to span a range in right ascension.     The order in which the fields are
observed was determined primarily by the desire to prevent overlapping
epochs of {\it HST} observations.   

{\it  JWST observability:}   Each of the selected Frontier Fields
positions was run through a preliminary {\it JWST} scheduling software and
confirmed to have extended {\it JWST} visibility periods. 
 \\
 \\

\begin{deluxetable*}{llllccccl}
\label{table1}
\tabletypesize{\footnotesize}
\tablecolumns{9}
\tablewidth{0pt}
\tablecaption{The Frontier Fields Locations }
\tablehead{
\colhead{Cluster}  &\multicolumn{2}{c}{Cluster Center (J2000)}  & \multicolumn{2}{c}{Parallel Center (J2000)} 
&\colhead{Epoch1} &\colhead{Epoch2} &\colhead{zodiacal $H_{F160W}$ \tablenotemark{a}} &\colhead{$E_{(B-V)}$ \tablenotemark{b}} \\
    &\colhead{$\alpha$}     & \colhead{$\delta$}       
&\colhead{$\alpha$}     & \colhead{$\delta$}  & {\it HST}  & {\it HST}   &\colhead{(AB mag/ $\sq\arcsec$)}  &      }
\startdata
Abell 2744          &  00:14:21.2  & -30:23:50.1  &00:13:53.6  & -30:22:54.3  &10/2013-12/2013  &5/2014-7/2014   &  22.2/21.9     & 0.012  \\  
MACSJ0416.1$-$2403  &  04:16:08.9  & -24:04:28.7  &04:16:33.1  & -24:06:48.7  &1/2014-2/2014 &7/2014-9/2014   &  22.4/22.3     & 0.036 \\
MACSJ0717.5$+$3745  &  07:17:34.0  & +37:44:49.0  &07:17:17.0  & +37:49:47.3  &9/2014-12/2014  &2/2015-3/2015   &  21.8/22.0    & 0.068 \\
MACSJ1149.5$+$2223  &  11:49:36.3  & +22:23:58.1  &11:49:40.5  & +22:18:02.3  &11/2014-1/2015  &4/2015-5/2015   &  21.9/22.0   & 0.020 \\
Abell S1063         &  22:48:44.4  & -44:31:48.5  &22:49:17.7  & -44:32:43.8  &10/2015-11/2015  &4/2016-6/2016   &  22.2/20.6    & 0.010  \\ 
Abell 370           &  02:39:52.9  & -01:34:36.5  &02:40:13.4  & -01:37:32.8  &12/2015-2/2016  &7/2016-9/2016   &  21.8/21.9     & 0.028 \\
\enddata
\tablenotetext{a}{Typical zodiacal background in $H_{F160W}$ for {\it HST} Epoch1 and Epoch 2 observations respectively; computed using {\it HST} exposure time calculator \\ and median observing date.}
\tablenotetext{b}{Schlafly \& Finkbeiner 2011, courtesy of the NASA/IPAC Extragalactic Database}
\end{deluxetable*}

\vspace{0.1 in}

\section{The Frontier Field Clusters\\ and Parallel Fields}

In February 2013,  the six Frontier Field clusters and their parallel
fields locations were finalized and announced prior to the {\it HST} Cycle
21 proposal deadline.     The Frontier Fields clusters are Abell 2744,
MACSJ0416.1-2403,  MACSJ0717.5+3745, MACSJ1149.5+2223,  Abell S1063
(also known as RXCJ2248.7-4431), and Abell 370  (Table 1).   These
clusters are at redshifts between 0.3 and 0.55,  and are among the most
massive known clusters at these redshifts (Table 2).  
All of the clusters had previous (shallow){\it HST} imaging,  with four clusters previously observed as part of the 
CLASH {\it HST} MCT survey (MACSJ0416.1-2403, MACSJ0717.5+3745, MACSJ1149.5+2223,  and Abell S1063) and all but Abell 370 were part of the MAssive Clusters Survey (Ebeling, Edge, \& Henry 2001).

\subsection{Abell 2744}
Abell 2744 is a massive X-ray luminous 
merging cluster at  $z=0.308$, (Couch \& Newell 1982; Abell, Corwin, \& Olowin 1989), 
also known as AC118 or ``Pandora's Cluster''.     It has a total X-ray
luminosity of $L_X = 3.1 \times 10^{45}$ erg s$^{-1}$ at $2-10$ keV
(Allen 1998), with  X-ray emission concentrated on the southern
compact core and extending to the northwest (Owers et al. 2011; Eckert et al. 2015).  
Its viral mass within the central 1.3 Mpc is $\sim 1.8 \times 10^{15} M_{\odot}$ (Merten
et al. 2011). The velocity dispersion is $\sigma = 1497 \pm 47$ km s$^{-1}$  (Owers et
al. 2011),  but shows two distinct structures, with the northern
substructure offset in velocity by $-1600$ km s$^{-1}$ and $\sigma \sim 800$
km s$^{-1}$ (Boschin et al. 2006;  Braglia et al.2007).   Abell 2744's complicated
velocity structure and lensing properties suggest that it is 
merging system with at least three separate sub-structures (Cypriano et
a. 2004; Braglia et al.2007;  Merten et al. 2011).  Weak lensing analysis by
Merten et al. (2011) identified four mass concentrations of core, N, NW, W of 2.2, 0.8, 1.1, 1.1 $\times 10^{14}$ $\Msun$ respectively,  with the NW structure showing evidence for
spatially separated dark matter, gas and galaxies.   Abell 2744 is also host to 
a powerful extended radio halo with $P_{1.4 GHz} = 1.5 \times 10^{25}$ W s$^{-1}$
(Giovannini, Tordi, \& Feretti 1999).

Despite its obviously complicated geometry,  Abell 2744 was one of the
strongest Frontier Field cluster candidates based on its lensing strength,  sky
location,  and pre-existing ancillary data. 
The pre-FF lensing model by Merten et
al. (2011; using the Zitrin et al. 2009 Light-Traces-Mass modeling method) found  
34 strong-lensed images of 11 galaxies in  {\it HST} F814W
imaging of the core of Abell 2744  (HST GO 11689, P.I.: R. Dupke),  
giving a core mass $\sim 2 \times 10^{14} M_{\odot}$.    This core region is $\sim 100\arcsec \times 100\arcsec$,  therefore fits within the {\it HST} {\it WFC3/IR} FOV of $2.2\arcmin \times 2.1\arcmin$. 
Analysis of preliminary models constructed by Zitrin and Richard separately
suggested a very high probability of magnifying a $z \sim 10$ galaxy to $H=27$ ABmag
within   {\it WFC3/IR} field of view. This high lensing
probability has been confirmed by subsequent models provided by the lensing map effort and independent teams 
(e.g. Coe et al. 2015; Atek et al. 2014; Zitrin et al. 2014;  Johnson et al. 2014; Lam et al. 2014; Richard et al. 2014;  Ishigaki et al. 2015; Wang et al. 2015;  Jauzac et al. 2015; Table 2).  

Abell 2744 has one of the darkest skies and lowest Galactic extinctions 
$E_{(B-V)} = 0.012$  (Schlafly \& Finkbeiner 2011)
of all the cluster candidates.  The typical zodiacal background in $H_{F160W}$ during the cluster IR epoch (10/2013-12/2013) and the parallel IR epoch (5/2014-7/2014) are $\sim 22.2$ and 21.9 ABmag per $\sq\arcsec$ respectively.  
At a declination of $-30$,  it is easily observable with {\it ALMA} and the {\it VLT}  but also within reach of Maunakea and the {\it Very Large Array}.    It has been extensively
studied by the {\it Chandra X-ray Observatory} (e.g. Kempner \& David 2004; Owers et al. 2011; Merten et al. 2011).  Abell 2744 was also observed during the {\it Spitzer} cryo-mission,  with MIPS 24 micron
and IRAC 3.6 - 8 micron observations (PI G. Rieke).   This cluster is part of the
Herschel Lensing Survey (Egami et al. 2010), with deep {\it Herschel Space Observatory} PACS 100/160
micron and SPIRE 250/350/500 micron imaging. 

The choice of parallel field was particularly challenging in this
case.   {\it HST} roll-angles with $>$ 30 day observing windows at both
orientations  placed the observable parallel
field either 6\arcmin east or west of the Abell 2744 core.      
However, the eastern parallel field location was undesirable because
of the presence of an unavoidable bright star.  Therefore the western
parallel field location ($\alpha$ = 00:13:53.6, $\delta$=-30:22:54.3, J2000) was chosen.   The parallel
field is $\sim 1-2\arcmin$ west of the NW and W sub-structures identified in Merten et
al. (2011).   The weak-lensing magnification boost from the cluster is
therefore predicted to significant,  with median magnification factors $\sim 1.14-1.21$ and maximum magnification factors $1.5-1.85$ for $1 < z < 9$ within the WFC3/IR pointing based on the pre-HFF v1.0 Merten model (Table 2). 

\subsection{MACSJ0416.1-2403}
MACSJ0416.1-2403 is a massive elongated
X-ray luminous cluster at z=0.397 (Ebeling et al. 2007; Ebeling et al. 2014) \footnote{This cluster's
  redshift is often incorrectly quoted as 0.42, based on preliminary analysis by
  Postman et al. 2012.}.     Its bolometric X-ray luminosity
is $L_x = 1.02 \times 10^{45}$ erg s$^{-1}$,  with a double-peaked profile
suggestive of a merging cluster (Mann \& Ebeling 2012).  The velocity dispersions for each
of these components are $\sigma$ = 779 $^{+22}_{-20}$ and 955 $^{+17}_{-22}$ (Jauzac et al. 2014;  Ebeling et al. 2014), and
the total mass enclosed within 950 kpc $\sim 1.2 \times 10^{15}  M_{\odot}$ (Jauzac et al. 2014; Grillo et al. 2015).  MACSJ0416.1-2403 was selected as one of 
five strong-lensing clusters for the {\it HST} MCT CLASH survey (Postman et al. 2012)  based on its large
Einstein radius ($\theta_E > 0.35 \arcsec$ at $z=2$). 
Prior to the Frontier Fields observations, Zitrin et al. (2013) found a high number of multiple images relative to its critical area in the
CLASH {\it HST} images,  likely due to its highly elongated and
irregular structure.    

Preliminary evaluation of MACSJ0416.1-2403's lensing  
models yielded moderate to high probabilities of detecting a $z \sim 10$ $H
\leq 27$ mag galaxy within the {\it WFC3/IR} field of view (Table 2). MACS0416.1-2403 is at a high ecliptic latitude with a Galactic
extinction E(B-V) = 0.036 (Schlafly \& Finkbeiner 2011).  The typical zodiacal background in $H_{F160W}$ during the cluster IR epoch (7/2014-9/2014) and the parallel IR epoch (1/2014-2/2014) are $\sim 22.3$ and 22.4 ABmag per $\sq\arcsec$ respectively.   At declination
$\sim$ -24,  this field is easily observable with {\it ALMA}, and also
available to  Maunakea.   A significant amount of data was
collected on this cluster as part of MACS and CLASH,  including shallow multi-band {\it HST} data, {\it Chandra} imaging, {\it Spitzer} warm-mission IRAC (PI Bouwens), and VLT spectroscopy (e.g. Grillo et al. 2015).  Additional Chandra imaging has since been obtained by C. Jones-Forman and S. Murray (Ogrean et al. 2015). However,  there are no legacy {\it Spitzer} cryogenic observations.    

MACSJ0416.1-2403 is notable for having a $J=10$, $V=13$ magnitude star
within 1\arcmin of the cluster core.   This star has a high proper
motion, with DSS and 2MASS imaging from the mid-1990s showing a
position a few arc-seconds north of its current (2014) {\it HST} {\it ACS} position. 
This star is included in the Frontier Fields {\it ACS} pointing, and lies
just off the {\it WFC3/IR} pointing,  resulting in scattered light and
saturated diffraction spikes in the Frontier Field images.   However,
this star is bright enough to act as an adaptive optics guide star,  
therefore provides a unique
opportunity to obtain AO imaging (e.g. Schrimer et al. 2015,  Gemini- GEMS)
and spectroscopy of the critical
curves surrounding a strong-lensing cluster.   

The MACSJ0416.1-2403 parallel field was chosen to lie westward of the
cluster pointing in order to avoid the bright eastern stars in the
{\it Spitzer} Frontier Field observations.   This orientation is
perpendicular to the elongation of the cluster on the sky, and
therefore we expect minimal contamination of the parallel field from the
cluster. The parallel field is predicted to have median magnification factors $\sim 1.09-1.16$ and maximum magnification factors $1.2-1.4$ for $1 < z < 9$ within the WFC3/IR pointing   based on the pre-HFF v1.0 Merten model (Table 2).

\subsection{MACS0717.5+3745}
MACSJ0717.5+3745 is an extremely massive  
X-ray luminous merging cluster at $z=0.545$ (Edge et al.2003).  
The X-ray luminosity between 0.1-2.4 keV is 
$3.3 \pm 0.2 \times 10^{45}$ km s${-1}$  (Edge et al. 2003).   The
cluster's velocity dispersion is $1660^{+120}_{-130}$ km s$^{-1}$ (Ebeling et al. 2007). 
Its optical and X-ray morphology shows a double peak and lack of 
center cluster core, with a filament towards extending southeast
  (Ebeling et al. 2004; Kartaltepe et al. 2008).   
This cluster also hosts the most power known radio source ($P(1.4 GHz) \sim 5 \times 10^{25} W Hz^{-1}$)  with a radio relic significantly offset from the cluster center to the north (van Weeren et al.2009). MACSJ0717.5+3745 was also chosen as one of the CLASH strong-lensing
clusters (Postman et al. 2012).  It has the largest known Einstein radius ($\sim$ 350 kpc,  Zitrin et al. 2009) and an estimated virial mass $\ge 2-3 \times 10^{15} M_{\odot}$ (Zitrin et al. 2009; Limousin et al. 2012).  Several pointings of {\it HST} {\it ACS} imaging were obtained previously 
by Ebeling in Cycle 12 (GO 9722).  Weak-lensing analyses of the pre-Frontier Fields {\it HST} imaging and ground-based Subaru imaging have confirmed the presence of the southeast filament, with a projected length $\sim$ 4.5 Mpc and true length of $\sim$ 18 Mpc (Jauzac et al. 2012;  Medenski et al. 2013)

Independent preliminary lensing models from Zitrin and Richard ranked
MACS0717.5+3745 as the strongest lenser of all the
considered clusters (see Table 2).
However,  MACSJ0717.5+3745 has the highest zodiacal background of all the Frontier Fields, as well as a relatively high Galactic extinction $E_{(B-V)} = 0.068$ (Schlafly \& Finkbeiner 2011). It has an ecliptic latitude of 15.4 degrees, 
with a typical zodiacal background in $H_{F160W}$ during the cluster IR epoch (2/2015-3/2015) and the parallel IR epoch (9/2014-12/2014) of $\sim 22.0$ and 
21.8 ABmag per $\sq\arcsec$ respectively.
It is also our northern-most cluster at declination $>30$,  placing it just out of reach of ALMA and other southern observatories.    As a CLASH
cluster,  significant shallow {\it HST} imaging,  ancillary wide-field
Subaru imaging, and a photometric redshift catalog are available.    
This cluster was also observed with {\it Spitzer} cryogenic mission with
both IRAC and MIPS (PI Kocevski) and the {\it Spitzer} warm-mission SURFSUP program (PI Bradac; Bradac et al. 2014),  as well as
with the {\it Herschel Space Observatory} (Egami et al. 2010).  A spectroscopic redshift catalog was recently published by Ebeling et al. (2014). 

The MACSJ0717.5+3745 parallel field was chosen to lie north-west of the 
cluster pointing in order to avoid the long cluster filament
extending to the south-east.  The parallel field is predicted to have median magnification factors $\sim 1.07-1.15$ and maximum magnification factors $1.17-1.42$ for $1 < z < 9$ within the WFC3/IR pointing based on the pre-HFF v1.0 Merten model (Table 2). 

\subsection{MACS1149.5+2223}
MACSJ1149.5+2223 at $z=0.543$ was discovered as part of the MACS survey as one of
the most X-ray luminous clusters known at $z>0.5$  (Ebeling et
al. 2001;  Ebeling et al. 2007).   Its 0.1-24 keV 
X-ray luminosity is $L_x = 1.76 \pm 0.04 \times 10^{45}$ erg s$^{-1} $
and it has a velocity dispersion $1840^{+120}_{-170}$ km s$^{-1}$
(Ebeling et al. 2007).     Its optically selected galaxy population
and X-ray morphology  is elongated within the cluster core,  but does
not show evidence of extended filaments (Kartaltepe et al. 2008). 
Spectroscopic studies and lensing analysis of previous {\it HST} {\it ACS} imaging (PI Ebeling; GO 9722) suggest four or more large-scale dark matter sub-haloes and a complex merger history  ( Zitrin \& Broadhurst 2009; see also Smith et al. 2009).  A CLASH strong-lensing
cluster  (Postman et al. 2012),   it has a large Einstein radius ($\sim$
170 kpc,  Zitrin \& Broadhurst 2009) and an estimated total mass $\sim 2.5
\times 10^{15} $ $M_{\odot}$ ( Zheng et al. 2012).    Based on the CLASH
imaging,   Zheng et al. (2012) reported a singly imaged z=9.6 galaxy candidate with
a magnification $\sim 14.5$ and observed $F160W$ magnitude $\sim 26.5$. 

Preliminary lensing models from Zitrin and Richard ranked
MACSJ1149.5+2223 as a moderate lenser (Table 2).
Its Galactic extinction is fairly low  $E(B-V) = 0.020$ (Schlafly \& Finkbeiner 2011), and a zodiacal background $\sim 22$ $H_{F160W}$ ABmag per $\sq\arcsec$ during the epochs of observation (cluster IR: 11/2014-1/2015;  parallel IR: 4/2015-5/2015). Initially, this cluster was not considered an ideal {\it HST} target as
only one bright guide star was known at the required orients.
However,  further investigation reveal a second guide star slightly
fainter than the nominal magnitude cut-off, and early
observations of MACSJ1149.5+2223 in Cycle 21 confirmed the suitability
of this guide star pair.  At  declination +22,  this cluster is barely
observable with ALMA but easily observed from Maunakea and other
northern observatories like the Very Large Array.   This cluster is
part of the Herschel Lensing Survey (Egami et al. 2010) and a GT Cycle 1 program (PI D. Lutz), and was targeted by {\it Spitzer} warm-mission SURFSUP IRAC imaging program (Bradac et al. 2014). 

The southern position for the MACSJ1149.5+2223 parallel field was
chosen to avoid a particularly bright star at the northern position.
 The parallel field is predicted to have median magnification factors $\sim 1.02-1.07$ and maximum magnification factors $1.1-1.3$ for $1 < z < 9$ within the WFC3/IR pointing
 based on the pre-HFF v1.0 Merten lensing model (Table 2).

\subsection{Abell S1063}
Abell S1063 (also known as RXC J2248.7-4431 and SPT-CL J2248-4431),  is the southern-most Frontier Fields cluster with $z=0.3461$ (Abell, Corwin, \& Olowin 1989; B\"{o}hringer et al. 2004; G\'{o}mez et al. 2012).    Abell S1063 is a massive cluster with a large velocity dispersion $1840^{+ 230}_{-150}$  km s$^{-1}$.   It's X-ray luminosity
between 0.5-2.0 keV  is $1.8\pm0.2 \times 10^{45}$ erg s$^{-1}$
(Williamson et al. 2011), and the cluster has one of the hottest known X-ray temperatures ($> 11.5$ keV) (G\'{o}mez et al. 2012).   
It is also among the strongest
Sunyaev-Zel'dovich (SZ) detected clusters in the South Pole Telescope
survey (Williamson et al. 2011),  with a SZ-derived mass $M_{500} \sim 1.4
\times 10^{15}$ $M_{\odot}$.   Like the other Frontier Field clusters,  the cluster galaxy
density map shows significant substructure, with an X-ray peak offset
from the primary galaxy density peak (G\'{o}mez et al. 2012).  Weak
lensing analysis also identified multiple substructures, and gives 
a mass of the central cluster in agreement with X-ray and SZ calculations
(Gruen et al.2013).    Selected as a CLASH cluster,  the {\it HST} imaging
revealed a quintuply lensed $z \sim 6$ galaxy (Monna et al. 2013, Balestra et
al. 2013).   The Herschel Lensing Survey (Egami et al. 2010) 
images show an associated 870 micron source,  one of the
highest redshift lensed sub-mm galaxies known (Boone et al.2013).

Abell S1063 is one of the less powerful lensers (Table 2) and most relaxed of the selected Frontier Fields clusters.   However, it is located in one of the darkest regions of
the sky, with a Galactic extinction of  $E_{(B-V)} = 0.010$ (Schlafly \& Finkbeiner 2011). The typical zodiacal background is 20.6 and 22.2  $H_{F160W}$ AB mag per $\sq\arcsec$ during the cluster IR epoch (4/2016-6-2016) and the parallel IR epoch (10/2015-11/2015) respectively.  
It is inaccessible with Maunakea but easily observed by ALMA and the VLT. 
As a SPT and CLASH cluster,  it had extensive spectroscopic and
ancillary data  already, including  shallow Chandra imaging (PI Romer), {\it Herschel} (Egami et al. 2010; also Open Time Cycle 2 program, PI T. Rawle),  SZ,  {\it Spitzer} cryo-mission MIPS and IRAC (PI G. Rieke),  and VLT spectroscopy (e.g. Balestra et al. 2013). 
Recently, Abell S1063 has been targeted by VLT MUSE integral field spectrograph (Karman et al. 2015). 

The Abell S1063 parallel field was chosen to the east of the cluster,
to avoid scattered light from the western bright stars in the {\it Spitzer}
and {\it HST} observations.    We note that  Gruen et al. (2013) report a
east-north-east cluster substructure which lies northward of the
AbellS1063 parallel field location.  The parallel field is predicted to have median magnification factors $\sim 1.02$ and maximum magnification factors $1.27-1.43$ for $1 < z < 9$ within the WFC3/IR pointing based on the pre-HFF v1.0 Merten lensing model.

\subsection{Abell370}
Abell370 (Abell 1958) at $z=0.375$ (Struble \& Rood
1999) is the host of the first known gravitational Einstein ring (Soucail et
al. 1987;  Paczynski 1987) and thus one of the best studied
strong-lensing clusters (e.g. Kneib et al. 1993; Smail et al.1996;
Bezecourt 1999a, b;  Broadhurst et al. 2008; Richard et al. 2010;
Medezinski et al. 2010;  Umetsu et al. 2011).  
Its total velocity dispersion is $\sim$ 1170 km s$^{-1}$ (Dressler al. 1999),  with the two main sub-structures showing internal velocity dispersions $\sim$ 850 km s$^{-1}$ (Kneib et al. 1993).  Abell 370's total bolometric X-ray luminosity is $L_x = 1.1 \times 10^{45}$ erg s$^{-1}$(Morandi, Ettori, \& Moscardini
2007).     X-ray, SZ, and lensing analyses of Abell 370 consistently yield a viral mass $\sim 1 \times 10^{15} M_{\odot}$  (e.g. Umetsu et al. 2011;  Richard et
al. 2010;  Morandi et al. 2007).   With {\it HST} {\it ACS} images taken shortly after the last {\it HST} refurbishment,   Richard et al. (2010) found significant offsets
between the peak X-ray emission and peaks of the lensing mass
distribution, and concluded that Abell 370 is likely the recent
merger of two equal mass clusters along the line of sight.   Like Abell 2744,  Abell 370 was not part of the the CLASH {\it HST} MCT survey (Postman et al. 2012). 

Abell 370 is one of the stronger lensers among the selected Frontier
Fields clusters,  with current models predicting $P(z=9.6) \sim 0.9$ (Table 2).  The typical zodiacal background is 21.9 and 21.8  $H_{F160W}$ AB mag per $\sq\arcsec$ during the cluster IR epoch (7/2016-9/2016) and the parallel IR epoch (12/2015-2/2016) respectively.   It has a Galactic foreground extinction $E_{(B-V)} = 0.028$, and is accessible with both Northern and Southern telescopes.   Abell 370 also has a rich legacy of archival data,  including Chandra imaging (PI Garmire), {\it Herschel} data from the PACS Evolutionary Probe (Lutz et al. 2011) and Herschel Multi-tiered Extragalactic Survey (Oliver et al. 2012), and cryogenic {\it Spitzer} data in the four IRAC  channels, IRS, and  MIPS (PIs Fazio; Rieke; Houck; Lutz; Dowell).  

For Abell 370, we choose the south-eastern parallel position in order to avoid
multiple bright stars north-west of the cluster and a possible
extension of cluster members to the north (Broadhurst et al. 2008).
The  parallel field is predicted to have the strongest weak-lensing boost, with median magnification factors $\sim 1.2-1.32$ and maximum magnification factors $1.35-1.63$ for $1 < z < 9$ within the WFC3/IR pointing based on the pre-HFF v1.0 Merten lensing model.

\subsection{Other Cluster Candidates}
We considered a number of potential Frontier Field clusters,
many of which are known to be exceptional lensers.    We excluded
Abell 1689,  Abell 1703, and the Bullet Cluster because of their low redshifts/ large
angular sizes of their critical curves relative to the {\it WFC3/IR} field
of view.   Abell 2537,  MACSJ1206.2-0747, MACSJ2129.4-0741,
MACSJ2214.9-1359, RCS2-2327.4-04,  RXJ1347.5-1144 all have low
ecliptic latitudes, and therefore have unacceptably high zodiacal backgrounds.    
 MACSJ0329.6-0211, MACSJ451.0+0006, MACSJ0520.7-1328,
MACSJ0744.9+3927  have high Galactic extinctions (E(B-V) $> 0.05$).
MACSJ0647.7+7015  and MACSJ744.9+3927 have numerous unavoidable bright stars in
the field.  MACSJ0647.7+7015, MACSJ744.9+3927, and MACSJ1423.8+2404 are unsuitable for deep ALMA
observations.  MACSJ0358.8-2995  has a foreground z=0.17 Abell cluster and very limited {\it HST} visibility. 
MACSJ0454.1-0300 is a weaker lenser with a moderate zodiacal background.
MACSJ0257-2325 had limited public ancillary data at the time of selection.  Additionally, these 
last three clusters are close in right ascension to each other and the UDF/GOODS-South field and MACSJ0416.1-2403, and therefore would have posed scheduling issues for {\it HST} over the course of the next several {\it HST} cycles.

\begin{deluxetable*}{lcccccccc}
\label{table2}
\tabletypesize{\footnotesize}
\tablecolumns{9}
\tablewidth{0pt}
\tablecaption{Frontier Fields: Cluster Properties and Ancillary Data}
\tablehead{
\colhead{Cluster} & \colhead{$z$ \tablenotemark{a}}  & \colhead{$M_{vir}$ \tablenotemark{a}}  &\colhead{$L_X$ \tablenotemark{a}}  &\colhead{P($z=9.6$) \tablenotemark{b}} & \colhead {Parallel $\mu$ \tablenotemark{c}} &\colhead{\it{Spitzer}}  &\colhead{ \it{Herschel} \tablenotemark{d}} & \colhead{ALMA \tablenotemark{e}}\\
                     &                             & \colhead{$M_{\odot}$}  &\colhead{erg s$^{-1}$}  &  $H \leq 27$  &    &\colhead{MIPS 24$\mu$m}   & \colhead{PACS/SPIRE}  &  }
\startdata
Abell 2744           &  0.308     & $1.8 \times 10^{15}$   & $3.1 \times 10^{45}$ &  0.69 $\pm$ 0.07    &1.14-1.21  & yes & 100/250/350/500  & yes\\  
MACSJ0416.1$-$2403   &  0.396     & $1.2  \times 10^{15}$      & $1.0 \times 10^{45}$   & 0.63 $\pm$ 0.12    &1.09-1.16  & no  & 100/250/350/500 & yes \\
MACSJ0717.5$+$3745   &  0.545     & $2-3  \times 10^{15}$      & $3.3 \times 10^{45}$   & 0.84 $\pm$ 0.05     &1.07-1.42  & yes &100/250/350/500  &no\\
MACSJ1149.5$+$2223   &  0.543     & $2.5  \times 10^{15} $      & $1.8 \times 10^{45}$   & 0.60 $\pm$ 0.10    &1.02-1.07  &no     &70/100/250/350/500  & yes\\
Abell S1063          &  0.348     & $1.4  \times 10^{15}$      & $1.8 \times 10^{45}$   & 0.69 $\pm$ 0.08      &1.02       &yes  &70/100/250/350/500  &yes \\
Abell 370           &  0.375      & $\sim 1  \times 10^{15}$        & $1.1 \times 10^{45}$  & 0.90 $\pm$ 0.08     &1.2-1.3    &yes  &100/250/350/500 &yes\\
\enddata
\tablenotetext{a}{See text for references for each cluster.}
\tablenotetext{b}{Median probability of lensing a $z=9.6$ background galaxy to apparent $H_{F160W}$ ABmag $\leq$ 27 within the WFC3/IR FOV, calculated using the pre-HFF v1.0 lensing models. }
\tablenotetext{c}{Median magnification factor $\mu$ in the parallel fields within the WFC3/IR FOV; based on the weak-lensing estimates from pre-HFF v1.0 Merten models.  Note that magnification factors may be larger at locations closer to the cluster.}
\tablenotetext{d}{See Rawle et al. 2016 for summary of {\it Herschel} and {\it Spitzer} cryogenic observations.  Note that the Herschel SPIRE 250/350/500 mm field of view covers both cluster and parallel fields for all but MACSJ0416.1-2403.}
\tablenotetext{e}{Visibility from ALMA}
\end{deluxetable*}

\begin{deluxetable}{lcc}
\label{table3}
\tabletypesize{\footnotesize}
\tablecolumns{3}
\tablewidth{0pt}
\tablecaption{Frontier Fields Observational Depths}
\tablehead{
\colhead{Camera/Filter} & \colhead{Exposure Time \tablenotemark{a}}  & \colhead{5 $\sigma$ \tablenotemark{b}} }
\startdata
{\it HST} {\it ACS/WFC} F435W          &    45 ks    &   28.8  \\
{\it HST} {\it ACS/WFC} F606W          &    25 ks             &   28.8 \\
{\it HST} {\it ACS/WFC} F814W          &   105 ks              &   29.1 \\
{\it HST} {\it WFC3/IR} F105W  &     60 ks              &   28.9    \\
{\it HST} {\it WFC3/IR} F125W  &    30 ks               &   28.6    \\
{\it HST} {\it WFC3/IR} F140W  &    25 ks               &   28.6     \\
{\it HST} {\it WFC3/IR} F160W  &    60 ks               &   28.7      \\
{\it Spitzer IRAC} 3.6$\mu$m &    50 ks          &   26.5       \\
{\it Spitzer IRAC} 4.5$\mu$m &    50 ks         &    26.0\\
\enddata
\tablenotetext{a}{Assuming 2500s per {\it HST} orbit.  {\it Spitzer} depths include previous archival observations.}
\tablenotetext{b}{Calculated for a point source within a 0.4\arcsec diameter aperture for {\it HST}.} 
\end{deluxetable}

\begin{figure*}
\plotone{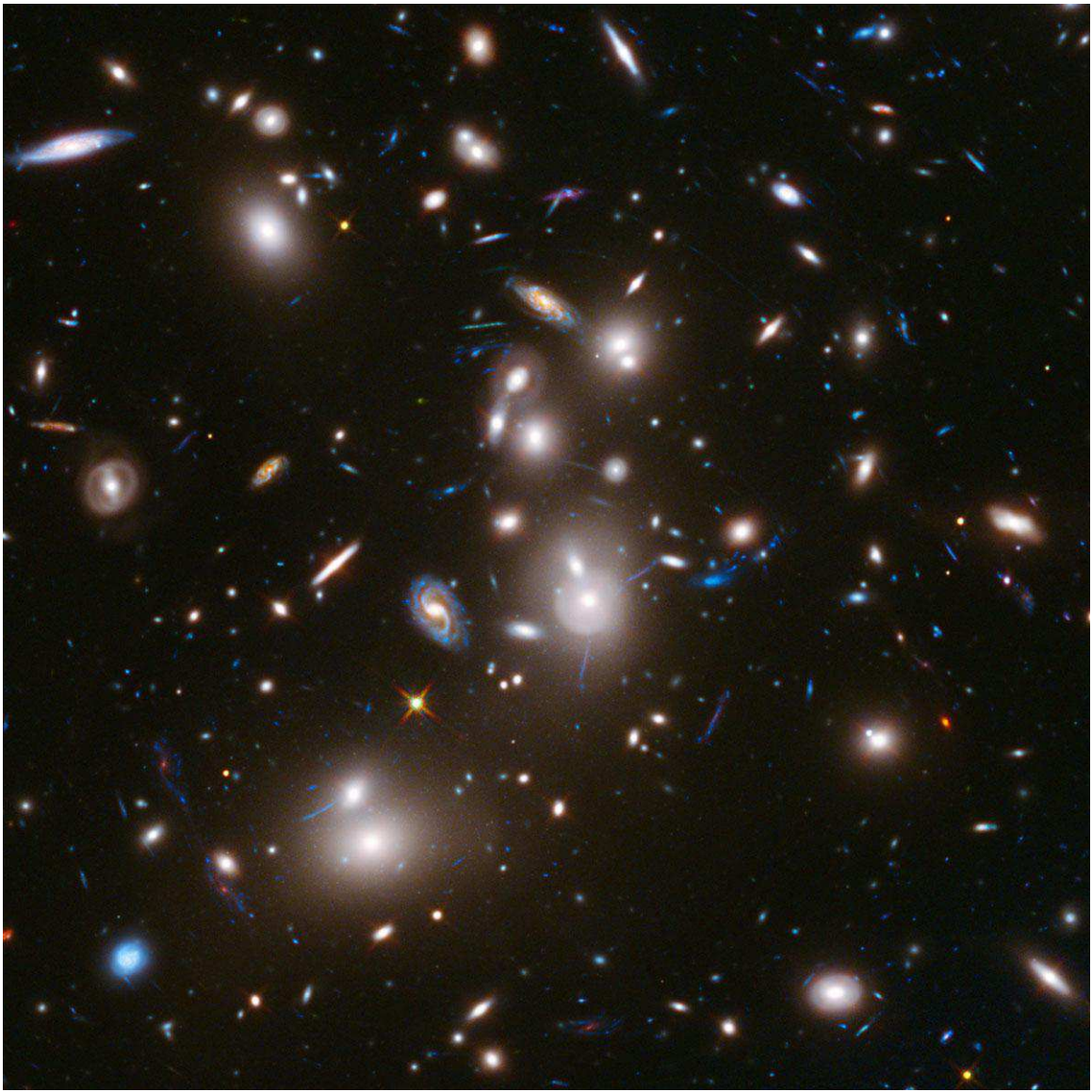}
\caption{{\it HST} full-depth image of Abell 2744, the first Frontier Field strong-lensing cluster. The central 1.5\arcmin $\times$ 1.5\arcmin is shown. }
\end{figure*}

\begin{figure*}
\plottwo{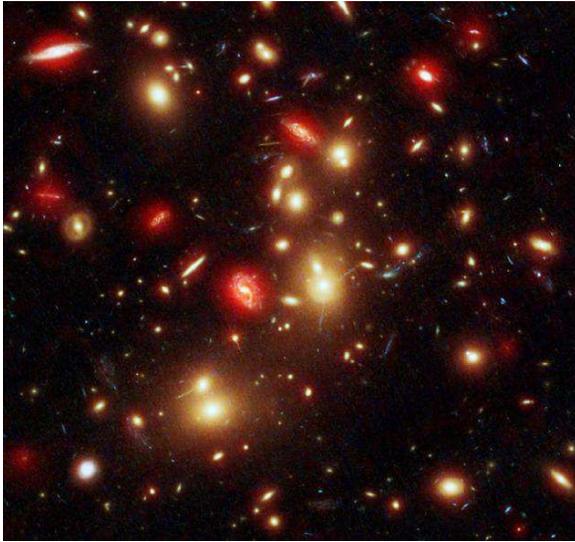}{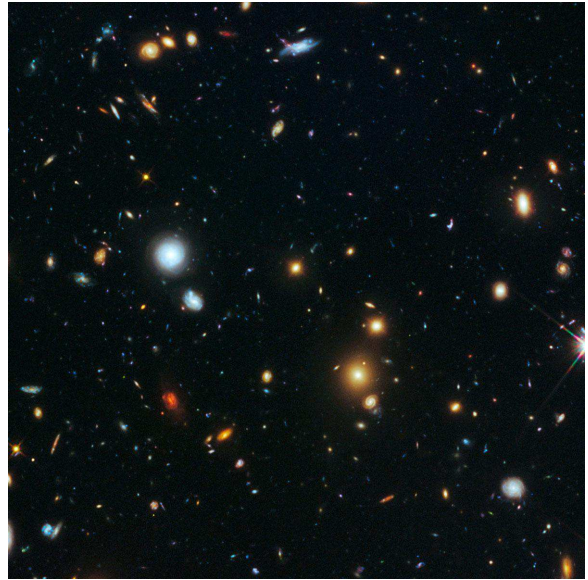}
\caption{{\it HST} $H_{F160W}$ $+$ {\it Spitzer IRAC} 3.6 and 4.6 micron image of Abell 2744 (left) and the {\it HST} full-depth image of Abell 2744 parallel field (central 1.5\arcmin $\times$ 1.5\arcmin).}
\end{figure*}

\begin{figure*}
\plottwo{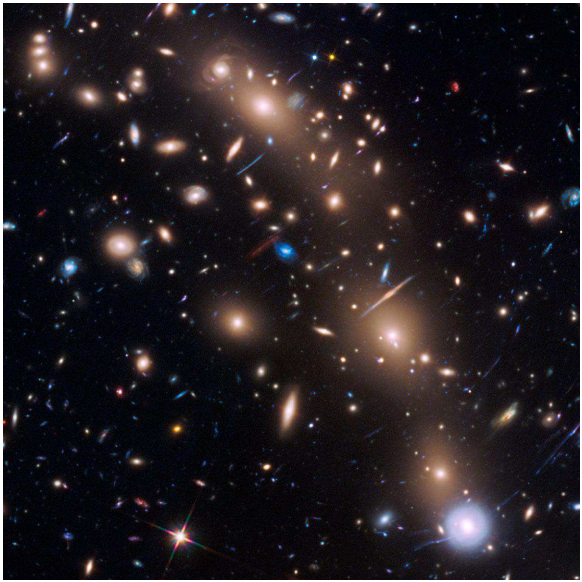}{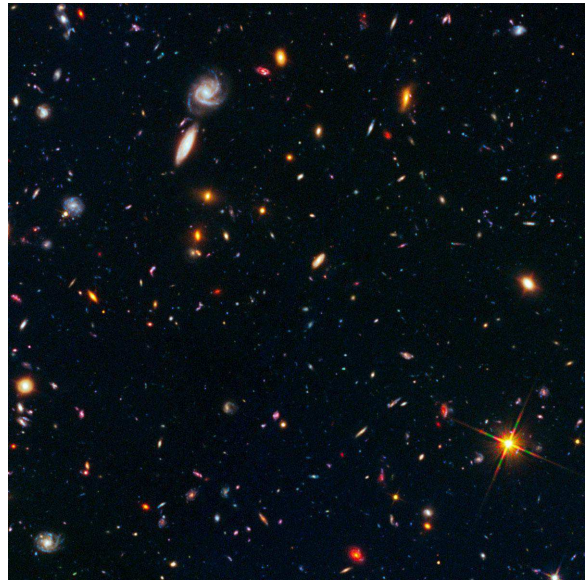}
\caption{{\it HST} full-depth image of MACSJ0416.1-2403 and its parallel field (central 1.5\arcmin $\times$ 1.5\arcmin)}
\end{figure*}

\begin{figure*}
\plottwo{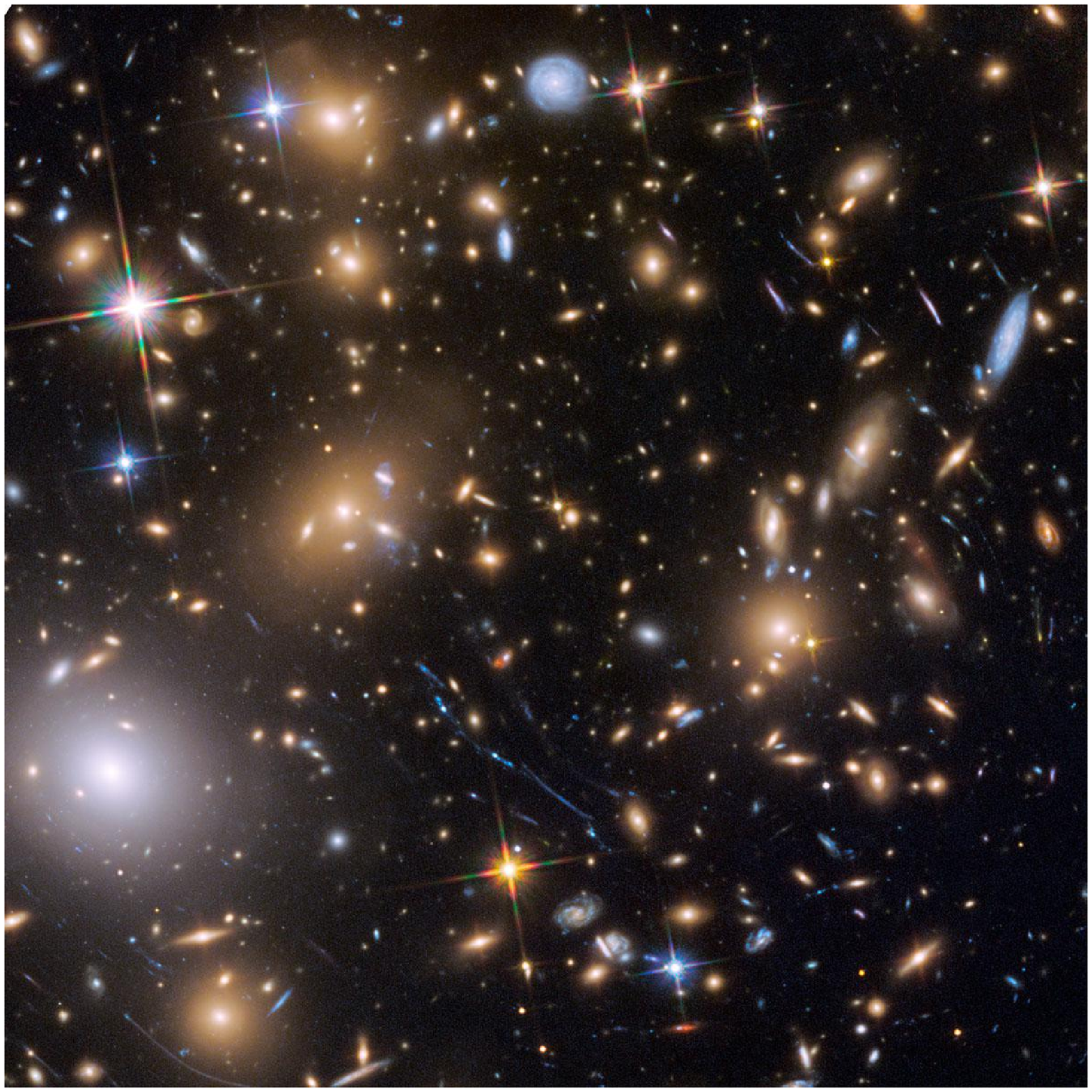}{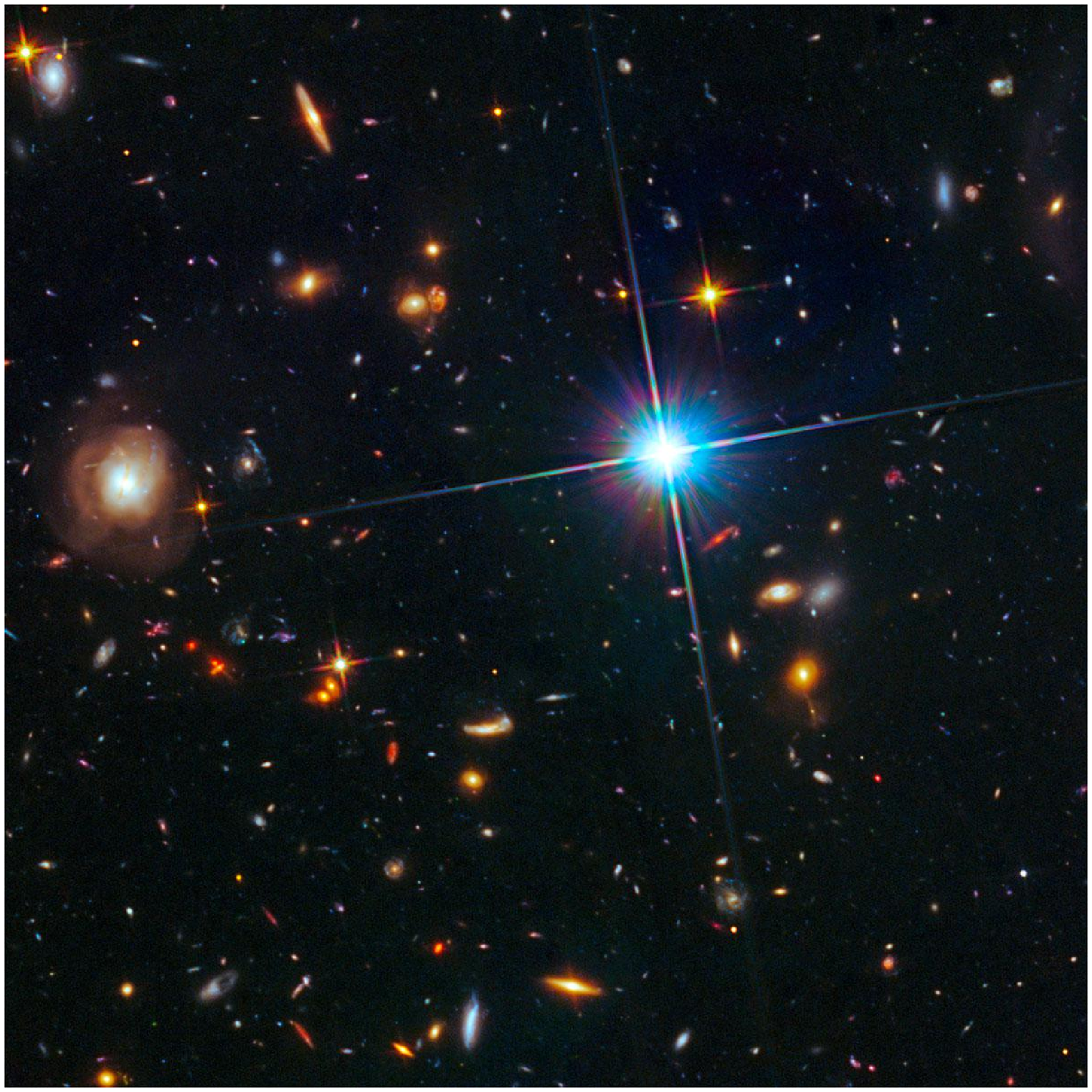}
\caption{{\it HST} full-depth image of MACSJ0717.5+3745 and its parallel field (central 1.5\arcmin $\times$ 1.5\arcmin)}
\end{figure*}

\begin{figure*}
\plottwo{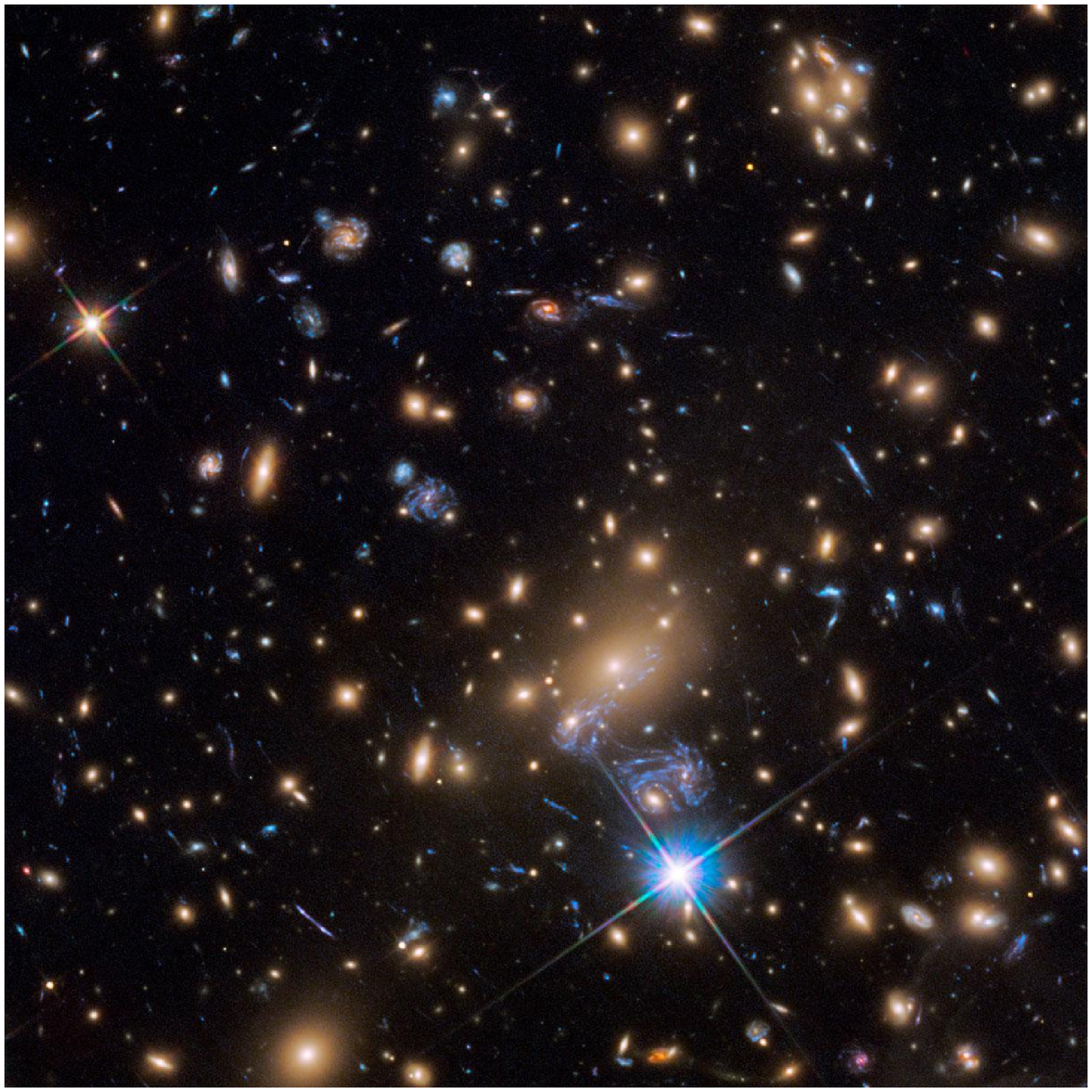}{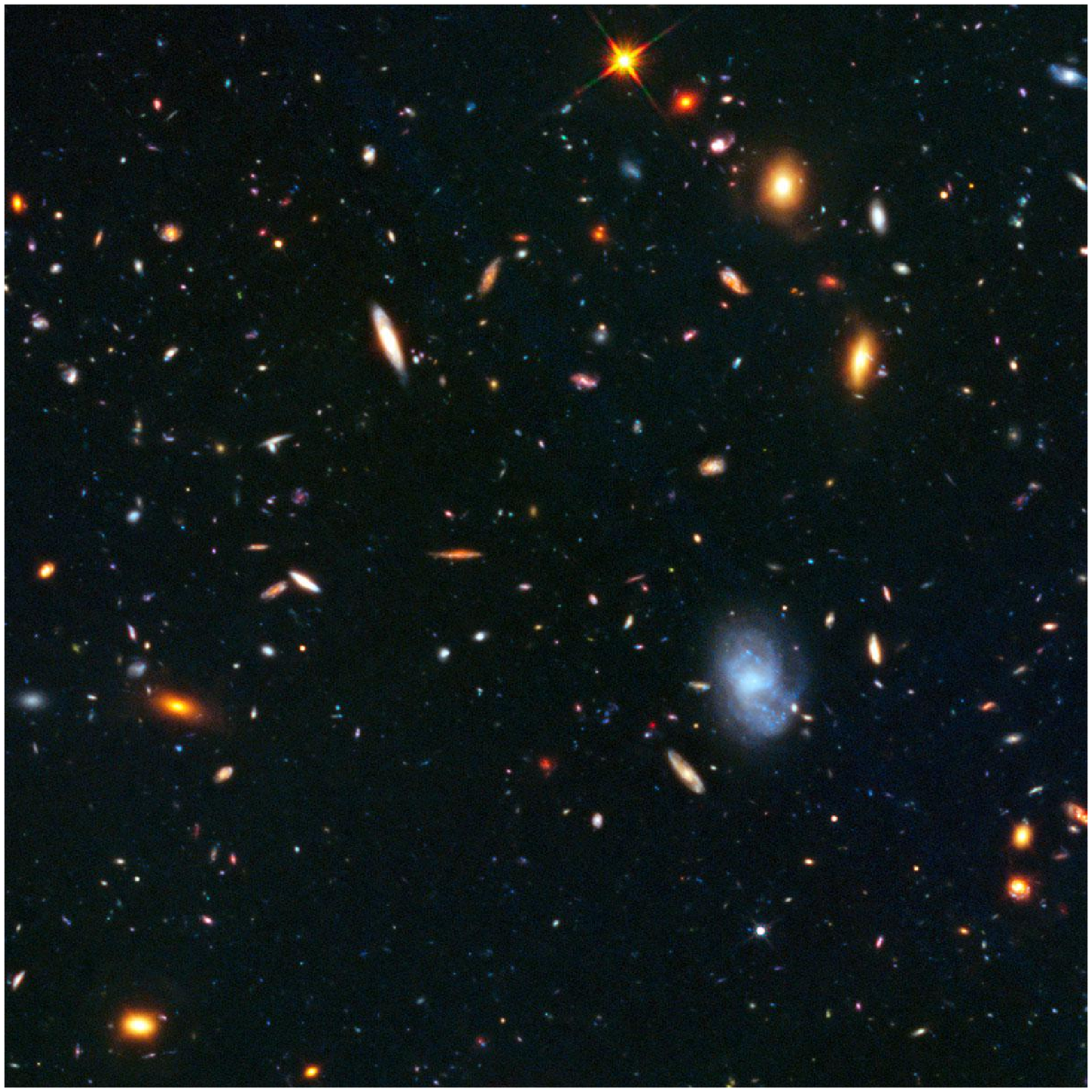}
\caption{{\it HST} full-depth image of MACSJ1149.5+2223 and its parallel field (central 1.5\arcmin $\times$ 1.5\arcmin)}
\end{figure*}

\section{Observations}
Deep optical and near-infrared imaging achieving $\sim$ 29th AB
magnitude 5$\sigma$ depths in seven {\it HST} bandpasses (ACS/WFC $B_{F435W}$, $V_{F606W}$, $I_{F814W}$, 
WFC3/IR $Y_{F105W}$, $J_{F125W}$, $JH_{F140W}$, $H_{F160W}$), from
0.4-1.6 microns are used to identify high-redshift galaxies ($z > 4$) using the
Lyman break drop-out technique (Table 3).   Deep {\it Spitzer} IRAC imaging at 3.6 and
4.5 microns place additional constraints on galaxy redshifts (Table 3).
Spectral energy distribution fitting of the multi-wavelength
photometry from the combined {\it HST} and {\it Spitzer} imaging (e.g. Merlin et al. 2016)
provide photometric redshifts, and estimates of the galaxy stellar masses and
recent star-formation histories (e.g. Castellano et al. 2016).    

The Frontier Field cluster observations have the same exposure times
as the parallel fields,  and similar observed depths.
However, the intrinsic depths for background galaxies lensed by the clusters
are deeper than the parallel fields (modulo the
contribution to the foreground by the cluster ICL and galaxies; see Livermore, Finkelstein, \& Lotz 2016; Merlin et al. 2016 for ICL subtraction strategies), with typical magnifications
across the cluster pointings  $\sim 1.5-2$  and small areas magnified by
factors as large as $>10-100$. 

\subsection{HST Observing Strategy}
Both the {\it Wide Field Camera 3} and {\it Advanced Camera for
Surveys} are used in concert at fixed {\it HST} roll-angles to probe each
Frontier Field cluster and a parallel `blank' field pair.    
Based upon the recommended depths and filter sets from the HDFI SWG
report,  we obtain 70 orbits per camera at a given roll angle,
for a total of 140 orbits per pointing for both the cluster and
parallel field.   The first four sets of Frontier Fields were awarded
DD time in Cycles 21 and 22 for a total of 560 orbits. Two more
Frontier Fields were approved for {\it Spitzer} DD observations in {\it Spitzer}
Cycle 11 and were awarded an additional DD 280 orbits in {\it HST} Cycle 23
after an external mid-term review of the program. \footnote{\url{www.stsci.edu/hst/campaigns/frontierfields/documents/FF_MidTermReview.pdf}}

{\it Filter Selection and Depths:}
The {\it ACS/WFC} observations are taken in the $B_{F435W}$, $V_{F606W}$, 
and $I_{F814W}$ filters, and the {\it WFC3/IR} observations are obtained in 
$Y_{F105W}$, $J_{F125W}$,  $JH_{F140W}$, and $H_{F160W}$ 
for both the parallel and cluster fields.   
The HDFI SWG recommended the $JH_{F140W}$ filter only for the cluster
pointings. This filter is most needed for discriminating between
$z \sim 9$ and higher redshift candidates, and it was felt that these would
be unlikely to be detected in the parallel fields.  However, subsequent
input from the community and discovery of bright $z \geq 9$ candidates 
resulted in the addition of the $JH_{F140W}$ to the
parallel field observations.  The number of orbits per filter/camera and estimated 
depths for 5 $\sigma$ point source measured within a 0.4$\arcsec$ diameter aperture
are given in Table 3.  

{\it Observational Cadence:}
Given the large number of orbits required for a given field and
orient,  we selected clusters for which {\it HST} observing windows 
of at least 30 days at fixed orients with suitable guide stars available at both orients.   
The HDFI SWG did not recommend dividing the observations of a given
field over multiple epochs to search for supernovae or other transient
objects.   Therefore,  the majority of data for each field was
obtained in two epochs of  $\sim$30-60 days (one for each camera/orient) separated by six months.
However, for those fields for which no pre-existing {\it HST} data was
available,  we obtained 1 advance visit with $ACS$/$I_{F814W}$ and/or 1 advance visit with
WFC3/$H_{F160W}$ to provide a template for transients and preliminary
catalogs for ground-based and {\it Spitzer} ancillary observations.  

During each main epoch of observations,  the {\it HST} {\it WFC3/IR} filter complement
was initially rotated through $Y_{F105W}$/$J_{F125W}$/$JH_{F140W}$/$H_{F160W}$ 
with a single filter per two-orbit
visit to facilitate the detection of high redshift
supernovae.   Our first set of  observations of Abell 2744 were impacted by
time-variable background in the {\it WFC3/IR} $Y_{F105W}$.  This is due to a known
HeI emission line at 10830\AA\ from the Earth's atmosphere,  which is detected by
{\it HST} when it observes at low limb angles at start or end of an orbit
and {\it HST}  is not in Earth's shadow.  During the course of our observations,  it was
determined that we could predict the times of highly variable sky
based upon the observational ephemeris (Brammer et al. 2014).    Therefore,  a subset of
our visits for MACSJ0416.1-2403 were changed to four-orbit visits,  with four half-orbit $Y_{F105W}$ exposures paired with four half-orbit $H_{F160W}$ exposures 
taken at start (or end) of each orbit when HeI emission was expected to have the largest impact.    
We found this strategy to work well for mitigating the impact of time-variable sky on the $Y_{F105W}$; remaining signatures of this effect, as well as time-variability in all IR filters when observing close to the bright Earth limb, are removed from our reduced data
using a modified IR ramp fitting algorithm (Robberto 2014; Hilbert 2014).  

Initially, the {\it ACS/WFC} filter complement was rotated through
$B_{F435W}$/$V_{F606W}$/$I_{F814W}$ throughout each
observing epoch as well.   At low sky backgrounds, {\it ACS/WFC} images are degraded by charge transfer efficiency (CTE) trails. While CTE trails from sources and hot pixels are now corrected in the standard pipeline, this correction is never perfect and results in residual noise above the ETC estimates.   However,  in reducing the observations for the first epoch of Abell 2744,  we found 
that the final combined images were greatly enhanced when ``self-calibrated'' to remove the
signature of trails in the darks and other detector-related
sources of noise\footnote{\url{www.stsci.edu/hst/acs/software/Selfcal}} \footnote{\url{blogs.stsci.edu/hstff/2013/05/24/calibration-is-in-the-works/}} (also Ogaz, Avila, \& Hilbert 2015).
Transient hot pixels in the darks are the major source of
this noise.  The imperfectly corrected hot pixels end up generating the same pattern of
residuals in all the images.  With multiple exposures ($> 8$), it is possible to self-calibrate out this pattern and regain $\sim$ 20\% in $B_{F435W}$ depth.  {\it ACS} undergoes a monthly annealing process in order to reduce the population of hot pixels. The structure of hot pixels in the darks are reset after the anneal, making the ’self-calibration’ software procedure less effective.  Therefore,  for later epochs of
observations, we grouped the {\it ACS/WFC}  $B_{F435W}$ and $V_{F606W}$ exposures in order
to straddle the planned {\it ACS} anneals.  The total number of $I_{F814W}$ exposures
is large enough to be self-calibrated with the number of  images taken
on  either side of the {\it ACS} anneals, and so they are interlaced with the $B_{F435W}$ 
and $V_{F606W}$ observations. 

{\it Dither Pattern:}
To maximize the sensitivity of the HST Frontier
Fields, especially toward the edges where strong
magnification is predicted, each epoch of observations
is constrained to a fixed HST roll angle with small
dithers between exposures.  The fixed roll-angle
requirement means that every HST visit within
an epoch is fine-guiding on the same pair of stars,
and therefore inter-visit dithering is highly effective.

To mitigate self-persistence between visits, we used 
an inter-visit dither pattern that displaced
any given two visits by $>1$ {\it WFC3/IR} pixel ($\sim$ 0.13\arcsec)
while still retaining overall compactness. This was achieved
by generating 35 pseudo-random dither locations from a
2D Sobol sequence covering a 6-pixel square.  At the same
time, pixel-phase dithering was achieved by modulating
this 6 pixel pattern by a secondary 35-element 2D Sobol sequence
sampling over pixel phase.  Pairings of ACS and WFC3 filters
were carefully matched to visit-specific dither locations
such that no filter had a pile-up of exposures in either absolute
location or in pixel phase.

The {\it HST} dithering within the Frontier Fields visits, comprising four half-orbit
exposures per filter, used the standard WFC3/IR "IR-DITHER-BLOB".
This intra-visit dither pattern had several
attractive features, including: good intra-visit subpixel
phase sampling for {\it WFC3/IR}; stepping across {\it WFC3/IR}
"blobs" of reduced detector sensitivity; and stepping across
the {\it ACS/WFC} CCD gap.  This intra-visit dither pattern is also
sufficient to reject cosmic ray impacts marring the four
half-orbit {\it ACS} exposures.  Because of the compactness
of IR-DITHER-BLOB, we do not completely fill in in the WFC3/IR ``deathstar''  
- a $\sim$ 6'' circular region of bad pixels - nor do we dither over the {\it WFC3/IR} ``wagon-wheel'',  an extended region on the right edge of the detector with low quantum efficiency and color-dependent structure which is not corrected by the existing flat fields.

{\it {\it WFC3/IR} Persistence:}
Like other sensitive {\it HST} {\it WFC3/IR} programs,  the Frontier Field observations are
scheduled to minimize the impact of IR detector persistence from bright objects previously observed by other {\it HST} programs (e.g. ``bad-actors''; Long, Baggett \&  MacKenty 2013).     Every Frontier Field
exposure is visually inspected for data quality issues including
persistence,  and additional checks for persistence are done \footnote{\url{archive.stsci.edu/prepds/persist}}.  
Most persistence impact small regions of the detector
and decays rapidly enough to effect only a few exposures, 
 and therefore can be effectively masked out in the final stacked
 {\it WFC3/IR} images.    However,  early {\it WFC3/IR} observations of the MACS0416.1-2403 parallel field were severely impacted by scanned {\it WFC3/IR} grism observations of 
 a bright star, for which persistence over $\sim$ 30\% of the {\it WFC3/IR} detector was visible 
 for $> 24$ hours after the grism observations (Long et al. 2014).   {\it HST} schedulers quickly responded to
change the following week's schedule to prevent repeating this
sequence of programs.   We triggered an {\it HST} Observation Problem Report (HOPR) 
to re-observe 10 orbits, 
and our input resulted in a change in the {\it HST} scheduling systems for the 
time buffer after such bad actors. 
Additional HOPR were called in Cycle 23 to repeat  persistence-effected observations for
Abell S1063 (8 orbits)  and Abell 370  (6 orbits). 

\subsection{HST Data Reduction}
We briefly describe here the Frontier Fields {\it HST} data pipeline and
resulting high-level science products.   For more details about the
{\it HST}  Frontier Fields data reduction,  please see Koekemoer et
al. 2016 (in prep) and the data release readme files associated with each {\it HST} dataset. 

Every incoming exposure is visually inspected and flagged for
artifacts, including satellite trails and asteroids,  IR persistence, and IR time-variable sky 
within a few days of acquisition.   Intermediate v0.5 stacked and drizzled images products are
produced with standard archival retrievals,  at 30 and 60 mas pixel
scales, with major artifacts masked.   The images are aligned with astrometric solutions 
based on previous {\it HST} and ground-based catalogs, initially compiled during the construction of the public Frontier Fields
lensing models in summer 2013.   Thus all MAST-hosted Frontier Fields lensing models and {\it HST} data products are aligned to the same astrometric grid.   

The v1.0 ``best effort'' image products are released within several weeks of the completion of the observing epoch for each cluster/parallel field pair at a given orient and camera configuration.  
These best effort image products includes the following  improvements above the v0.5 releases: 

$\bullet$ reprocessing of all exposures using the most recent {\it ACS} and {\it WFC3} 
calibration files (darks, flats, biases).

$\bullet$ improved astrometric alignment between filters, and cameras

$\bullet$ improved treatment of {\it ACS/WFC} bias destriping 

$\bullet$ ``self-calibration'' applied to the {\it ACS/WFC} images to remove
residual detector noise/artifacts, including correction for CTE in the
darks  

$\bullet$ masking of any new {\it WFC3/IR} ``blobs'' and additional
persistence sources

$\bullet$ correction for {\it WFC3/IR} time-variable sky in the
ramp-fitting, which most strongly affects the F105W observations due
to the HeI emission but also impacts all IR filters when observing close to the bright Earth limb.

$\bullet$ inclusion of {\it HST} imaging from other programs in the same filters in the stacked images to achieve maximum depths.

\subsection{Spitzer Observations}

In {\it Spitzer} Cycles 9, 10, and 11, all six Frontier Fields clusters were observed with IRAC channels 1 and 2 (3.6 and 4.5 micron) with Director's Discretionary time.   Combined with archival data, the final images are expected to have nominal 5-sigma point source sensitivities of 26.6 AB mag at 3.6 microns and 26.0 AB mag at 4.5 microns. However, contributions from confusion and the intra-cluster light may mean the observations are less sensitive at the cluster core.  Two of the clusters (MACS0717.5+3745 and MACS1149.5+2223) are in a previously approved {\it Spitzer} Cycle-9 program SURFSUP (PI M. Bradac, 90009), and two of the clusters (MACS0416.1-2403 and MACS0717.1.5+3745) were observed by the Cycle-8 program iCLASH (PI R. Bouwens, 80168).  Due to conflicting roll angle constraints with {\it HST} and {\it Spitzer}, the IRAC and {\it HST} fields of view could not be matched in position angle.   Furthermore, to maximize the depth of these observations the observing windows were constrained to the epochs with the lowest background. As a result there are significant "flanking field" areas covered by IRAC to 25h depth around the main {\it HST} fields.   For the reduced {\it Spitzer} data products, readme files, and additional information, please see Capak et al. (in prep) and  
\url{irsa.ipac.caltech.edu/data/SPITZER/Frontier/}

\section{Lensing Models \& Predictions}
In order to enable study of background lensed galaxies by a broad 
cross-section of the extra-galactic community,  
the {\it HST} Frontier Fields team has also supported the development and public release of lensing maps for each selected cluster.   
The initial lensing models were based on data taken {\it before} the Frontier Fields observing campaign to ensure that the community
could make use of the Frontier Fields data as soon as possible (Table 4). Five independent teams  (Brad\u{a}c; Clusters As TelescopeS, PI Kneib \& Natarajan; Zitrin \& Merten; Sharon; Williams), using a diversity of approaches (Brad\u{a}c et al. 2005; LENSTOOL: Julio \& Kneib 2009; Zitrin et al. 2009; Merten et al. 2009; GRALE: Mohammed et al. 2014),  coordinated to adopt the same input  archival {\it HST} and ground-based datasets, the same redshifts, 
and multiple image identifications.  These models were made public on MAST prior to the HST Frontier Fields observations in autumn 2013\footnote{\url{www.archive.stsci.edu/prepds/frontier/lensmodels/}}.   The initial pre-FF model predictions for the galaxy numbers and volumes probed at 
high-redshift are described in Coe, Bradley, \& Zitrin (2015). 

However, these first pre-FF models have been rapidly superseded.  The deep 
{\it HST} data have resulted in an unprecedented set of strong-lensed arcs and multiple images for constraining the cluster potentials (e.g. Lam et al. 2014; Jauzac et al. 2015; Wang et al. 2015; Kawamata et al. 2016; Jauzac et al. 2014; Diego et al. 2015).  Subsequent observations with the GLASS {\it HST} {\it WFC3/IR} grism GO program (Treu et al. 2015; Schmidt et al. 2014; see \url{archive.stsci.edu/prepds/glass/}) 
and new ground-based spectroscopic campaigns have greatly increased the number and accuracy of the redshifts for the background lensed FF galaxies
(Wang et al. 2015, Hoag et al. 2015; Johnson et al. 2014; Richard et al. 2014; Ebeling et al. 2014; Grillo et al. 2015; Balestra et al. 2015).  The detection of 
a lensed SNIa in MACSJ0416.1-2403 has also provided a strong constraint on its true magnification (Rodney et al. 2015).  The discovery of the multiply-imaged SN Refsdal in MACSJ1149.5+2223  (Kelly et al. 2014)  sparked an independent coordinated effort to predict the time delays and re-appearance of this supernovae in another image of the host galaxy (Treu et al. 2016; Kelly et al. 2016; also Rodney et al. 2016).   Additional programs have sought to understand and improve the systematics inherent in the different modeling approach (e.g. Zitrin et al. 2015; Mohammed et al. 2016; Harvey, Kneib, \& Jauzac 2016; Meneghetti et al., in prep.).  

The Frontier Fields lensing models will continue to be refined, as the HST Frontier Fields observing program 
proceeds through September 2016, new ancillary spectroscopic and weak-lensing datasets are acquired, and the modeling methods improve.   This investment is critical for ensuring the Frontier Fields' legacy for {\it JWST} studies.  
To continue to provide the best models to the broader community,  a renewed effort to  update the existing lensing models and incorporate new FF and ancillary data began in May 2015 for Abell 2744 and MACS0416.1-2403 (Table 4).   The resulting models were publicly released in autumn 2015.    
A second round of lensing coordination is set to begin in summer 2016, and will encompass the last four clusters.  The delivery of MACSJ1149.5+2223 and MACSJ0717.5+3647 models are due in February 2017, with final delivery of  Abell S1063 and Abell 370 models due in February 2018. 

\section{Summary}
We present the motivation and survey design for the Frontier Fields, a Director's Discretionary time program with {\it HST} and {\it Spitzer} to see deeper into the distant universe than ever before.  Six strong-lensing clusters and
six parallel fields are observed, probing galaxies to observed optical/near-infrared magnitudes of $\sim$ 29 ABmag and $10-100$ times fainter in regions of high magnification.  We explain the primary scientific goals of the Frontier Fields,  the selection criteria for the fields,  and the detailed properties of each Frontier Field cluster and parallel.   We describe the {\it HST} and {\it Spitzer} observing programs, and the coordinated Frontier Fields lensing model effort.

The {\it HST} Frontier Fields observations of the last cluster (Abell 370) and its parallel field will complete in September 2016, and the coordinated lensing models will be updated in 2017-2018. The full {\it Spitzer} Frontier Fields observations are complete and were publicly released in early 2016.  The first Frontier Fields observations have already probed galaxies during the epoch of reionization to intrinsic luminosities fainter than any previously seen (e.g. Livermore, Finkelstein, \& Lotz 2016; Castellano et al. 2016; Atek et al. 2015; Laporte  et al. 2015;  Zitrin et al. 2014). The full dataset will place strong statistical constraints on the faint end of the luminosity function during this era (Robertson et al. 2015).  At the time of publication of this article, over 70 refereed publications and 3 conferences have been devoted to or based in part on the Frontier Fields. These works include studies of high-redshift galaxies in the cluster and parallel fields; new cluster lensing models and dark matter maps;  supernovae/transient studies; intra-cluster light and cluster evolution studies; and ancillary observations probing highly-lensed background sources with major ground-based facilities.  These data and associated models will provide a unique legacy for future high-redshift universe studies with the {\it James Webb Space Telescope}. 

The Frontier Fields program was initiated by STScI Director Dr. Matt Mountain using Director’s Discretionary Time on the Hubble Space Telescope. We wish to acknowledge the Hubble Deep Fields Initiative science working group members for conceiving and recommending the Frontier Fields program: J. Bullock (chair), M. Dickinson, S. Finkelstein, A. Fontana, A. Hornschemeier Cardiff, J. Lotz, P. Natarajan, A. Pope, B. Robertson, B. Siana, J. Tumlinson, and M. Wood-Vasey.  We also thank the mid-term Frontier Fields review committee for their service: J Bullock, M. Dickinson, R. Ellis, M. Kriek, S. Oey, S. Seitz, S. A. Stanford, and J. Tumlinson.  We recognize the contributors to the current Frontier Field lensing models: M. Brada\u{c}, 
S. Allen, D. Applegate, B. Cain, A. Hoag, P. Kelly, P. Schneider,  
T. Schrabback, T. Treu, A. von der Linden, J.-P. Kneib, P. Natarajan, H. Ebeling, J. Richard, B. Clement, M. Jauzac, E. Jullo,  
M. Limousin, E. Egami, J. Merten,  A. Zitrin, I. Balestra, 
M. Bartelmann, N. Benitez, A. Biviano, T. Broadhurst, 
M. Carrasco, D. Coe, N. Czakon, M. Donahue, T. Eichner, R. Ellis, 
C. Giocoli, S. Golwala, C. Grillo, O. Host, L. Infante, S. Jouvel, 
D. Lemze, A. Mercurio, E. Medezinski, P. Melchior, 
A. Molino, M. Meneghetti, A. Monna, J. Moustakas, L. Moustakas, 
T. Mroczkowski, M. Nonino, M. Okabe, M. Postman, J. Rhodes, 
P. Rosati, J. Sayers, S. Seitz, K. Umetsu, K. Sharon, T. Johnson,  
M. Bayliss, L. Wiliams, I. Mohammed, P. Saha, J. Liesenborgs,  
K. Sebesta, M. Ishigaki, R. Kawamata, M. Oguri, J. M. Diego,  
D. Lam, and J. Lim.   Finally, we thank David Adler, George Chapman, Bill Workman, Ian Jordan, Alan Welty,  Karen Levay, Scott Fleming, Brandon Lawton, Carol Christian, Tony Darnell, Frank Summers, Kathy Cordes, Bonnie Eisenhamer, Lisa Frattare, Ann Jenkins,  Hussein Jirdeh,  John Maple, Holly Ryer,  Ray Villard,  Tracy Vogel, and Donna Weaver for their contributions to the {\it HST} Frontier Fields effort.

Based on observations obtained with the NASA/ESA {\it Hubble Space Telescope}, retrieved from the {\it Mikulski Archive for Space Telescopes} (MAST) at the Space Telescope Science Institute (STScI). STScI is operated by the Association of Universities for Research in Astronomy, Inc. under NASA contract NAS 5-26555. This work is based in part on observations made with the {\it Spitzer Space Telescope}, which is operated by the Jet Propulsion Laboratory, California Institute of Technology under a contract with NASA.  This work utilizes gravitational lensing models produced by PIs Bradač, Natarajan \& Kneib (CATS), Merten \& Zitrin, Sharon, Williams, and the GLAFIC and Diego groups. This lens modeling was partially funded by the {\it HST} Frontier Fields program conducted by STScI.

\clearpage
\begin{deluxetable}{llcclcccccc}
\tabletypesize{\footnotesize}
\tablecolumns{11}
\tablewidth{0pt}
\tablecaption{ Frontier Field Lensing Models \tablenotemark{a}}
\tablehead{
\colhead{Team} & \colhead{Method}  & \colhead{Parallel?}  &\colhead{Version}  &\colhead{Data\tablenotemark{b}}  &\colhead{Abell 2744}  &\colhead{MACSJ0416.1} &\colhead{MACSJ0717} &\colhead{MACSJ1149.5}  &\colhead{Abell S1063}  &\colhead{Abell 370} }
\startdata
CATS  & LENSTOOL  & no   & 1   & pre-HFF  & 10/2013   & 12/2013  & 12/2013  & 12/2013  & 12/2013  & 12/2013\\
          &       &      & 2   & HFF-     &  $-$      & 10/2014    & $-$      & $-$              & $-$       & $-$            \\
          &       &      & 2.1 & HFF-     & 9/2015    & $-$        & $-$      & $-$              & $-$       & $-$            \\
          &       &      & 2.2 & HFF-     & 9/2015    & $-$        & $-$      & $-$              & $-$       & $-$            \\
          &       &      & 3   & HFF+     & 9/2015    & 9/2015     & $-$      & $-$              & $-$       & $-$            \\
          &       &      & 3.1 & HFF+     & 9/2015    & 9/2015     & $-$      & $-$              & $-$       & $-$            \\
\\
\hline
\\
Sharon  & LENSTOOL  & no  & 1    & pre-HFF  & 10/2013 & 12/2013  & 12/2013  & 12/2013  & 12/2013  & 12/2013\\
        &           &     & 2    & pre-HFF  & 5/2014  & 5/2014   & $-$        & $-$       & $-$       & $-$            \\
        &           &     & 3    & HFF+     & 9/2015  & 9/2015   & $-$        & $-$      & $-$       & $-$            \\
\\
\hline
\\
Zitrin    & NFW   & no    &1   & pre-HFF  & 9/2013 & 9/2013  & 9/2013  & 9/2013  & 9/2013  & 9/2013\\
 \&Merten & LTM   & no    &1   & pre-HFF  & 9/2013 & 9/2013  & 9/2013  & 9/2013  & 9/2013  & 9/2013\\
          & LTM-G & no    &1   & pre-HFF  & 9/2013 & 9/2013  & 9/2013  & 9/2013  & 9/2013  & 9/2013\\         
          & WL    &yes    &1   & pre-HFF  & 9/2013 & 9/2013  & 9/2013  & 9/2013  & 9/2013  & 9/2013\\
          & NFW   & no    &3   & HFF+     & 9/2015 & 9/2015  & $-$     & $-$      & $-$       & $-$            \\
          & LTM-G & no    &3   & HFF+     & 9/2015 & 9/2015  & $-$     & $-$       & $-$       & $-$            \\
\\
\hline
\\
GLAFIC  &         &no   &1    &HFF-   & 11/2014     &  $-$     & $-$          & $-$              & $-$              & $-$             \\
        &         &     &3    &HFF+   &  2/2016     &  2/2016  & $-$              & $-$              & $-$       & $-$            \\
\\
\hline
\\
Williams & GRALE  &no  & 1   &pre-HFF  & 9/2013 & 9/2013     & 9/2013  & 9/2013  & 9/2013  & 9/2013\\
         &        &    & 2   & HFF-    &  $-$    & 10/2014    & $-$      & $-$      & $-$       & $-$            \\
         &        &    & 3   & HFF+    &  $-$    & 11/2015    & $-$      & $-$      & $-$       & $-$            \\
         &        &    & 3.1 & HFF+    & 11/2015 & 11/2015    & $-$      & $-$      & $-$       & $-$            \\
\\
\hline
\\
Brad\u{a}c    &    &yes   & 1       &pre-HFF  & 9/2013 & 9/2013  & 9/2013  & 9/2013  & 9/2013  & 9/2013\\
              &    &      &2        &HFF-     & 9/2015 & $-$     & $-$    & $-$      & $-$       & $-$            \\
\\
\hline
\\
Diego     &        &?      & 3        & HFF+      &  $-$      & 2/2016    & $-$              & $-$              & $-$       & $-$            \\ 
\enddata
\tablenotetext{a}{See \url{archive.stsci.edu/prepds/frontier/lensmodels/} for models; \url{www.stsci.edu/hst/campaigns/frontier-fields/Lensing-Models} for lensing primer and description of the different methods.}
\tablenotetext{b}{pre-HFF models were constructed prior to the FF observations with the coordinated input data; HFF- models were constructed with FF observations but without coordination between the teams; HFF+ models were constructed with FF observations with coordinated inputs between teams.}
\end{deluxetable}


\begin{references}
\reference{}Abell, G. 1958, ApJS, 3, 211
\reference{}Abell, G., Corwin, H.G., \& Olowin, R. 1989, ApJS, 70, 1
\reference{}Abraham, R., Tanvir, N.R., Santiago, B.X., Ellis, R.S., Glazebrook, K., \& van den Bergh, S. 1996, MNRAS, 279, 47
\reference{}Alavi, A. et al. 2014, ApJ, 780, 143
\reference{}Allen, S.W. 1998, MNRAS, 296, 392
\reference{}Atek, H. et al. 2014, ApJ, 786, 60
\reference{}Atek, H. et al. 2015, ApJ, 814, 69
\reference{}Balestra et al. 2013, A\&A, 559, L9
\reference{}Balestra et al. 2015, ApJS, in press, arXiv:1511.02522
\reference{}Barger, A.J., Cowie, L.L., Mushotzky, R.F., Yang, Y., Wang, W.-H., Steffen, A.T., \& Capak, P. 2005, AJ, 129, 578
\reference{}Beckwith, S. et al. 2006, AJ, 132, 1729
\reference{}Behroozi, P., Wechsler, R., \& Conroy, C. 2013, ApJ, 770, 57 
\reference{}Bezecourt, J., Kneib J.P., Soucail G., \& Ebbels T.M.D. 1999a, A\&A, 347, 21
\reference{}Bezecourt, J., Soucail G., Ellis R.S., Kneib J.-P. 1999b, A\&A 351, 433
\reference{}B\"{o}hringer, H. et al. 2004, A\&A, 425, 367 
\reference{}Boone, F. et al. 2013, A\&A, 559, L1
\reference{}Boschin, W., Giarardi, M., Spolaor, M., \& Barrena, R. 2006, A\&A, 449, 461
\reference{}Bouwens, R.J.  et al. 2003, ApJ, 595, 589
\reference{}Bouwens, R.J., Illingworth, G., Franx, M., \& Ford, H. 2007, ApJ, 670, 928
\reference{}Bouwens, R. et al. 2010, ApJ, 725, 1587
\reference{}Bouwens, R. et al. 2014, ApJ, 795,126
\reference{}Boylan-Kolchin, M., Bullock, J.S., \& Garrison-Kimmel, S. 2014, MNRAS, 443, L44
\reference{}Bradley, L.D. et al. 2012, ApJ, 760, 108
\reference{}Brad\u{a}c, M. et al. 2005, A\&A 437, 39
\reference{}Brad\u{a}c, M. et al. 2009, ApJ, 706, 1201
\reference{}Brad\u{a}c, M. et al. 2014, ApJ, 785, 108
\reference{}Braglia, F., Pierini, D., \& Bohringer, H. 2007, A\&A, 470, 425
\reference{}Braglia, F.G., Pierini, D., Biviano, A., \& B\"{o}hringer, H. 2009, A\&A, 500, 947
\reference{}Brammer, G. et al. 2013, ApJ, 765, 2
\reference{}Brammer, G.,  N. Pirzkal, P. McCullough, \& J. MacKenty 2014, Instrument Science Report WFC3, "Time-varying Excess Earth-glow
Backgrounds in the WFC3/IR Channel", \url{www.stsci.edu/hst/wfc3/documents/ISRs/WFC3-2014-03.pdf}
\reference{}Broadhurst, T., Umetsu, K., Medezinski, E., Oguri, M. \& Rephaeli, Y. 2008, ApJ, 685, L9 
\reference{}Casertano, S. et al. 2000,  AJ, 120, 2747
\reference{}Castellano, M. et al. 2016, A\&A submitted, arXiv:1603.02461
\reference{}Coe, D. et al. 2013,  ApJ, 762, 32
\reference{}Coe, D., Bradley, L., \& Zitrin A. 2015, ApJ, 800, 84
\reference{}Cohen, J. et al. 2000, ApJ, 538, 29
\reference{}Couch, W.J. \& Newell, E.B. 1982, PASP, 94, 610
\reference{}Cypriano, E.S., Sodre, L, Kneib, J.-P., \& Campusano, L.E. 2004, ApJ, 613, 95
\reference{}da Cunha, E. et al. 2013, ApJ, 765, 9
\reference{}Davis, M. et al. ApJL, 660, 1
\reference{}Diego, J.M. et al. 2005, MNRAS 360, 477
\reference{}Diego, J. M. et al. 2005, MNRAS, 362, 1247
\reference{}Diego, J. M. et al. 2007, MNRAS, 375, 958
\reference{}Diego, J. M. et al. 2015, MNRAS, 446, 683
\reference{}Dickinson, M. E. 1999, AIPC 470 122D,  ``A complete NICMOS map of the Hubble Deep Field North'', The 9th astrophysics conference: After the dark ages, when galaxies were young (the Universe at $2<Z<5$).
\reference{}Ebeling, H., Edge, A.C., \& Henry, J.P., 2001, ApJ, 553, 668
\reference{}Ebeling, H., Barrett, E., Donovan, D., Ma, C.-J., Edge, A.C., \& van Speybroeck, L. 2007, ApJ, 661, L33
\reference{}Ebeling, H., Ma, C.-J., \& Barrett, E. 2014, ApJS, 211, 21
\reference{}Eckert, D. et al. 2015, Nature, 528, 105
\reference{}Edge, A.C., Ebeling, H., Bremer, M., R{\"o}ttgering, van Haarlem, M. P., Rengelink, R., \& Courtney, N.J.D. 2003, MNRAS, 339, 913
\reference{}Egami, E. et al. 2010, A\&A, 518, L12
\reference{}Ellis, R. et al. 2013, ApJ, 763, 7
\reference{}Fan, X. et al. 2006, AJ, 132, 117
\reference{}Ferguson, H.C., Dickinson, M., \& Williams, R.E. 2000, ARA\&A, 38, 667
\reference{}Fern{\'a}ndez-Soto, Lanzetta, \& Yahil 1999
\reference{}Finkelstein, S., Papovich, C., Giavalisco, M., Reddy, N., Ferguson, H.C., Koekemoer, A., \& Dickinson, M. 2010, ApJ, 719, 1250
\reference{}Finkelstein, S., et al. 2012, ApJ, 756, 164
\reference{}Finkelstein, S., et al. 2015, ApJ, 810, 71
\reference{}Fontana, A. et al. 2003, ApJ, 594, 9
\reference{}Fontana, A. et al. 2006, A\&A, 459, 745
\reference{}Ford, H. et al. 1998, SPIE, 3356, 234
\reference{}Franx, M. 2003, {\it HST} GO 9723,  ``Deep NICMOS Imaging''
\reference{}Giacconi, R. et al. 2001, ApJ, 551, 624
\reference{}Giavalisco, M. et al. 2004, 600, 93
\reference{}Giovannini, G., Tordi, M. \& Feretti, L. 1999, New Astronomy, 4, 141
\reference{}G\'{o}mez, P.L. et al. 2012, AJ, 144, 79
\reference{}Graham,  et al. 2009
\reference{}Grazian et al. 2006, A\&A, 449, 951
\reference{}Grillo, C. et al. 2015, ApJ, 800, 38
\reference{}Grogin, N. et al. 2012, ApJS, 197, 35
\reference{}Gruen, D et al. 2013, MNRAS, 432, 1455
\reference{}Guhathakurta, P., Tyson, J.A., \& Majewski, S.R. 1990, ApJ, 357, 9
\reference{}Hilbert, B. 2014, Instrument Science Report WFC3 2014-17, "Updated non-linearity calibration method for WFC3/IR", \url{www.stsci.edu/hst/wfc3/documents/ISRs/WFC3-2014-17.pdf}
\reference{}Hinshaw, G. et al. 2013, ApJS, 208, 19
\reference{}Hoag, A. et al. 2016, ApJ, submitted, arXiv:1603.00505
\reference{}Hornschemeier, A. et al. 2000, ApJ, 541, 49
\reference{}Illingworth, G. \& Bouwens, R.J. 2010, AIPC, {\it THE FIRST STARS AND GALAXIES: CHALLENGES FOR THE NEXT DECADE}, volume 1294, 202
\reference{}Ishigaki, M., Kawamata, R., Ouchi, M., Oguri, M., Simasaku, K., \& Ono, Y. 2015, ApJ, 799, 12
\reference{}Jauzac, M. et al. 2012, MNRAS, 426, 3369
\reference{}Jauzac, M. et al. 2014, MNRAS, 443, 1549
\reference{}Jauzac, M. et al. 2015, MNRAS, 452, 1437
\reference{}Johnson, T.L., Sharon, K., Bayliss, M.B., Gladders, M., Coe, D., \& Ebeling, H. 2014, ApJ, 797, 48
\reference{}Jones, T.J. et al. AJ, 149, 107
\reference{}Julio, E. et al. 2007, NJPh, 9, 447
\reference{}Julio, E. \& Kneib, J.P. 2009, MNRAS, 395, 1319
\reference{}Karman, W. et al. 2015, A\&A 574, A11
\reference{}Kartaltepe, J., Ebeling, H., Ma, C.-J., \& Donovan, D. 2008, MNRAS, 389, 1240
\reference{}Kawamata, R.,  Oguri, M.,  Ishigaki, M.,  Shimasaku, K., \& Ouchi, M. 2016, 	ApJ, 819, 114
\reference{}Kneib, J.-P., Mellier, Y., Fort, B. \&  Mathez, G. 1993, A\&A, 273, 367
\reference{}Kneib, J.-P., \& Natarajan, P. 2011, ARA\&A, 19, 47
\reference{}Kelly, P. et al. 2015, Science, 347, 1123
\reference{}Koekemoer, A. et al. 2011, ApJS, 197, 36
\reference{}Labb{\'e}, I. et al. 2003, ApJ, 591, 95
\reference{}Labb{\'e}, I. et al. 2010, ApJ, 716, L103
\reference{}Labb{\'e}, I. et al. 2013, ApJ, 777, L19
\reference{}Lam, D. et al. 2014, ApJ, 797, 98
\reference{}Laporte, N. 2015, A\&A, 575, 92
\reference{}Lawrence, A. et al. 2007, MNRAS, 379, 1423
\reference{}Limousin, M. et al. 2012, A\&A, 554, A71
\reference{}Livermore, R. et al. 2012, MNRAS, 427, 688
\reference{}Livermore, R., Finkelstein, S., \& Lotz, J.M. 2016, ApJ submitted, arXiv:1604.06799
\reference{}Long,K.S., Baggett, S. \&  MacKenty, J. 2013, WFC3 Instrument Science Report 2013-07, 
"Characterizing Persistence in the WFC3 IR Channel: Observations of Omega Cen", \url{www.stsci.edu/hst/wfc3/documents/ISRs/WFC3-2013-07.pdf}
\reference{}Long, K.S., Baggett,S.M., MacKenty, J. \&  McCullough, P. 2014, WFC3 Instrument Science Report 2014-14, "Attempts to Mitigate Trapping Effects in Scanned Grism Observations of
Exoplanet Transits with WFC3/IR", \url{www.stsci.edu/hst/wfc3/documents/ISRs/WFC3-2014-14.pdf}
\reference{}Lowenthal, J. et al. 1997, ApJ  481, 673
\reference{}Lutz, D. et al. 2011, A\&A, 532, 90
\reference{}Madau, P., Ferguson, H.C., Dickinson, M.E., Giavalisco, M., Steidel, C., \& Fruchter, A. 1996, MNRAS, 283, 1388
\reference{}Mann, A. W.  \& Ebeling, H. 2012, MNRAS, 420, 2120
\reference{}McCullough, P.,  J. Mack, M. Dulude, \& B. Hilbert 2016, Instrument Science Report WFC3 2014-21, "Infrared Blobs: Time-dependent Flags", \url{www.stsci.edu/hst/wfc3/documents/ISRs/WFC3-2014-21.pdf}
\reference{}McLure, R.J. et al. 2011, MNRAS, 418, 2074
\reference{}McPartland, C., Ebeling, H., Roediger, E., \& Blumenthal, K. 2016, MNRAS, 455, 2994
\reference{}Medenski, E. et al. 2013, ApJ, 777, 42
\reference{}Menanteau, F. et al. 2012, ApJ, 748, 7
\reference{}Merlin, E. et al. 2016, A\&A submitted, arXiv:1603.02460
\reference{}Merten, J. et al. 2009, A\&A, 500, 681
\reference{}Merten, J. et al. 2011, MNRAS, 417, 333
\reference{}Mobasher, B. et al. 2004
\reference{}Mohammed, I.  Liesenborgs, J.,  Saha, P.,  Williams, L. L. R. 2014, MNRAS, 439, 2651
\reference{}Mohammed, I.,  Saha, P.,  Williams, L. L. R.,  Liesenborgs, J., \& Sebesta, K. 2016, MNRAS, in publication, arXiv:1507.01532
\reference{}Monna, A. et al. 2013, MNRAS, 438, 1417
\reference{}Montes, M. \& Trujillo, I. 2014, ApJ, 794, 137
\reference{}Oesch, P. et al. 2007, ApJ, 671, 1212
\reference{}Oesch, P. et al. 2010, ApJL, 709, 21
\reference{}Oesch, P. et al. 2012, ApJ, 745, 110
\reference{}Oesch, P. et al. 2013, ApJ, 772, 136
\reference{}Oesch, P. et al. 2013, ApJ, 773, 75
\reference{}Oesch, P. et al. 2016, ApJ, 819, 129
\reference{}Ogaz, S., Avila, R., \& Hilbert B. 2015, STScI Newsletter, volume 32,  \url{blogs.stsci.edu/newsletter/files/2015/03/FFCalibration.pdf}
\reference{}Ogrean, G. et al. 2015, ApJ, 812, 153
\reference{}Oguri, M. 2010, PASJ, 62, 1
\reference{}Oliver, S. et al. 2012, MNRAS, 424, 1614
\reference{}Ono, Y. et al. 2013, ApJ, 777, 155
\reference{}Owers, M.S. et al. 2011, ApJ, 728, 27
\reference{}Paczynski, B. 1987, Nature, 325, 572
\reference{}Papovich, C., Dickinson, M., Ferguson, H.C. 2001, ApJ, 559, 620
\reference{}Pirzkal, N. et al. 2013, ApJ, 775, 88
\reference{}Pope, A. et al. 2006, MNRAS, 370, 1185
\reference{}Pope, A. et al. 2016, ApJ, submitted
\reference{}Postman, M. et al. 2012, ApJS, 199, 25
\reference{}Rawle, T. et al. 2016, MNRAS, 459, 1626
\reference{}Richard, J., Kneib, J.-P., Limousin, M., Edge, A., \& Jullo, E. 2010, 
\reference{}Richard, J. et al. 2014, MNRAS, 444, 268
\reference{}Rix, H.-W., et al. 2004, ApJS, 152, 163
\reference{}Robberto, M. 2014, Proc. SPIE 9143, Space Telescopes and Instrumentation 2014: Optical, Infrared, and Millimeter Wave, eds. Oschmann, J., Clampin, M., Fazio, G., \& MacEwen, H. 
\reference{}Robertson, B., Ellis, R.S., Dunlop, J.S., McLure, R.J., Stark, D.P., \& McLeod D. 2014, ApJ, 796, 27
\reference{}Robertson, B., Ellis, R., Furlanetto, S.R., \& Dunlop, J. 2015, ApJ, 802, 19
\reference{}Rodney, S. et al. 2015, ApJ, 811, 70
\reference{}Rodney, S. et al. 2016, ApJ, 820, 50
\reference{}Schlafly, E. F. \& Finkbeiner, D. 2011, ApJ, 737, 103
\reference{}Schlegel, D., Finkbeiner, D., \& Davis M. 1998, ApJ, 500, 525
\reference{}Schmidt, K. et al. 2014, ApJ, 782, L36
\reference{}Schrimer, M., Carrasco, E.R., Garrel, V., Winge, C., Neichel, B., \& Vidal F. 2015, ApJS, 217, 33
\reference{}Scoville, N. et al. 2007, ApJS, 172, 38
\reference{}Smail, I.,  Dressler, A.,  Kneib, J.-P.,  Ellis, R. S., Couch, W. J., Sharples, R. M., \& Oemler, A. 1996, ApJ,  469,508
\reference{}Smith, G. et al. 2009, ApJ, 707, 163
\reference{}Spergel, D. et al. 2003, ApJS, 148, 175
\reference{}Songaila, A., Cowie, L.L., \& Lilly, S.J. 1990, ApJ,  348, 371
\reference{}Soucail, G., Fort, B., Mellier, Y. \& Picat, J. P. 1987, A\&A,  172, L14
\reference{}Steidel, C. et al. 1996, AJ, 112, 352
\reference{}Stiavelli, M. et al. 1999, A\&A 343, 25
\reference{}Struble, M.F. \& Rood, H.J. 1999, ApJS, 125, 35
\reference{}Thompson, R., Storrie-Lombari, L.J., Weymann, R.J., Rieke, M.J., Schneider, G., Stobie, E., \& Lytle, D. 1999, AJ, 117, 17
\reference{}Thompson, R. 2003, ApJ, 596, 748 
\reference{}Treister, E. et al. 2004, ApJ, 616, 123
\reference{}Treu, T. et al. 2015, ApJ, 812, 114
\reference{}Umetsu, K., Broadhurst, T.,  Zitrin, A.,  Medezinski, E.,  Coe,
D.,  \& Postman, M. 2011, ApJ, 738, 41
\reference{}van Weeren, R.J., R{\"o}ttgering, H.J., Br{\"u}ggen, \& Cohen, A. 2009, A\&A, 505, 991
\reference{}Wang, X., et al. 2015, ApJ, 811, 29
\reference{}Windhorst, R. et al.2011
\reference{}Williams, R. E. et al. 1996, AJ, 112, 1335
\reference{}Williams, R. E. et al. 2000, AJ, 120, 2735
\reference{}Williamson, R. et al. 2011, ApJ, 738, 139
\reference{}Yan, H. et al., 2011, ApJ, 728, 22
\reference{}Zheng, W. et al. 2012, Nature, 489, 406
\reference{}Zitrin, A., Broadhurst, T., Rephaeli, Y., Sadeh, S. 2009, ApJ, 707, L102 
\reference{}Zitrin, A., \& Broadhurst, T. 2009, ApJ, 703, L132
\reference{}Zitrin, A. et al. 2013, ApJ, 762, L30
\reference{}Zitrin, A., et al. 2014, ApJL, 793, 12
\reference{}Zitrin, A., et al. 2015, ApJ, 801, 442
\end{references}
\end{document}